\documentclass[aps,preprint,preprintnumbers,nofootinbib,showpacs,showkeys]{revtex4}

\usepackage{amsmath}
\usepackage{amssymb}
\usepackage{color}
\usepackage{dsfont}
\usepackage[english]{babel}
\usepackage{graphicx,accents}

\def\bea{\begin{eqnarray}}
\def\eea{\end{eqnarray}}
\def\lla{\left\langle}
\def\rra{\right\rangle}
\def\za{\alpha}
\def\zb{\beta}
\def\zc{\gamma}

\def\ssc{\scriptscriptstyle}
\def\lsim{\mathrel{\raise.3ex\hbox{$<$\kern-.75em\lower1ex\hbox{$\sim$}}} }
\def\gsim{\mathrel{\raise.3ex\hbox{$>$\kern-.75em\lower1ex\hbox{$\sim$}}} }

\makeatletter
\DeclareRobustCommand{\cev}[1]{%
  \mathpalette\do@cev{#1}%
}
\newcommand{\do@cev}[2]{%
  \fix@cev{#1}{+}%
  \reflectbox{$\m@th#1\vec{\reflectbox{$\fix@cev{#1}{-}\m@th#1#2\fix@cev{#1}{+}$}}$}%
  \fix@cev{#1}{-}%
}
\newcommand{\fix@cev}[2]{%
  \ifx#1\displaystyle
    \mkern#23mu
  \else
    \ifx#1\textstyle
      \mkern#23mu
    \else
      \ifx#1\scriptstyle
        \mkern#22mu
      \else
        \mkern#22mu
      \fi
    \fi
  \fi
}

\makeatother

\begin{document}


\title{Observables and Dynamics  Quantum to Classical from a Relativity Symmetry and Noncommutative-Geometric Perspective}

\author{Chuan Sheng Chew}

\author{Otto C. W. Kong}
\email{otto@phy.ncu.edu.tw}

\author{Jason Payne}

\address{Department of Physics and 
Center for High Energy and High Field Physics,\\
National Central University, Chung-Li 32054, Taiwan}



\vspace*{-.7in}
\begin{abstract}
With approaching quantum/noncommutative models for the deep 
microscopic spacetime in mind, and inspired by our recent picture
of the (projective) Hilbert space as the model of physical space behind
basic quantum mechanics, we reformulate here the WWGM formalism 
starting from the canonical coherent states and taking wavefunctions 
as expansion coefficients in terms of this basis. This provides us 
with a transparent and coherent story of simple quantum dynamics 
where both the  {\em wavefunctions} for the pure states {\em and 
operators acting on them} arise from \textit{the single space/algebra},
which exactly includes the WWGM observable algebra.
Altogether, putting the emphasis on building our theory out of the 
underlying relativity symmetry -- the centrally extended Galilean 
symmetry in the case at hand -- allows one to naturally derive both 
a kinematical and a dynamical description of a  quantum particle, 
which moreover recovers the corresponding classical picture (understood 
in terms of the Koopman-von Neumann formalism) in the appropriate 
(relativity symmetry contraction) limit. Our formulation here is 
the most natural framework directly connecting all of the relevant 
mathematical notions and we hope it may help a general physicist 
better visualize and appreciate the noncommutative-geometric
perspective behind quantum physics. 
It also helps to inspire and illustrate our 
perspective on looking at quantum mechanics and quantum 
physics in general in direct connection to the notion of quantum
(deformed) relativity symmetries and the corresponding 
quantum/noncommutative models of spacetime as various
levels of approximations all the way down to the Newtonian.
\end{abstract}



\maketitle

\section{Introduction to the Quantum Relativity Perspective}
Some years before the turn of the century,  the idea that physical spacetime 
should be modeled, at least at the deep microscopic scale, by some form 
of noncommutative geometry \cite{C} started to get more and more 
appreciation from physicists. A major part of it has to do with considerations 
about the compatibility of the basic notion is quantum mechanics and
a theory of geometrodynamics like Einstein general relativity \cite{DFK}.
Noncommutative geometry being obtainable from string theory \cite{SW}
also helps to promote the idea. The development of noncommutative
geometry in mathematics has been driven however in quite a part by
the study of operator algebras as motivated by the observable algebra  
of quantum mechanics. Quantum mechanics is about noncommutative
geometry though the observable algebra is usually taken as related
only to the quantum phase space. The Newtonian space is still assumed
to be the right model for the physical space. Yet, the position observable
which should be like coordinates in the space are given by elements
of the noncommutative observable algebra. In a theory of particle 
dynamics, the only truly physical notion about the space is really the
totally of all possible positions for a free particle. The Newtonian
space is the configuration space for a Newtonian free particle, and
the phase space is like a sum of two copies, as the momentum space
is isomorphic to the configuration space. In fact, one can start from 
the relativity symmetry group, the Galilean $G(3)$ group in the case,
and obtain the spaces from a representation of the latter, deriving
the dynamics as symmetry flows generated by the Hamiltonian as
the energy observable. For quantum mechanics, there is no notion
of configuration space and the phase space can only be obtained
from the so-called projective representation of $G(3)$, which 
really means a unitary representation of a bigger group, the $U(1)$ 
central extension $\tilde{G}(3)$. Our key perspective is that the
latter should be taken as the relativity symmetry behind quantum
mechanics and the phase space taken as the quantum model for
the physical space. In Ref.\cite{066}, we have given a brief report
on the key result that the perspective not only put the related 
notions of quantum mechanics on the same footing as those for
Newtonian mechanics, one can also retrieve the Newtonian
ones, including the Newtonian space, as classical limits of the
quantum notions. The current paper presents the full picture of
the dynamics. The nonexistence of any useful notion of the
quantum configuration space, or momentum space, here
is simply a direct parallel of the nonexistence of independent 
notion of space and time in the Einstein theory. The quantum 
phase space as the quantum physical space only splits into the 
configuration and momentum space at the Newtonian limit. 

Our background theoretical/mathematical setting is given by that of 
deformed special relativity \cite{dsr-o,dsr-o2,dsr-o3,dsr-o4,dsr-o5,dsr,dsr1,dsr2,dsr3,dsr4,dsr-l}, 
within a Lie algebra/group 
framework \cite{030,talk1,060,066}.  The key theme of deformed special 
relativity is to look at alternative theories of special relativity,
including possible candidates for deep microscopic physics with 
interesting features including noncommutativity, to which the Einstein
theory is an approximation. Note that the notion of spacetime
noncommutativity at the zero gravity limit has no conflict at all with 
necessity of it in the presence of gravity. Our formulation shows a 
noncommutative model for the physical space is already there behind 
basic quantum mechanics. It is essentially a deformation of the Newtonian, 
with deformation here taken on the level of the relativity symmetry. 
The latter exerts its effects to the spacetime model \cite{066} and every
aspects of the dynamical theory. We present, however, the inverse-deformation
as symmetry contractions \cite{060}. The fully deformed/stabilized level
has full noncommutativity \cite{030}, all without considering gravity.
The theory with gravity would be the `general quantum relativity', in which 
we expect the noncommutativity to be the quantum gravitational dynamics
 \cite{talk1,talk2}.   Results of ours then suggests any such models should 
probably include the case of quantum mechanics at some limit before 
getting to the Newtonian. 

A fundamental idea behind our approach lies in obtaining a fully 
grounded picture with a definite starting point of noncommutativity in 
physics, namely simple quantum mechanics and its commutative limit. The 
current paper actually presents only results relating to the latter part. The 
mathematics of  noncommutative geometry starts from a (noncommutative) 
algebra, and the algebra of observables for quantum mechanics with the 
Heisenberg structure is, of course, noncommutative. This noncommutativity 
is, however, only between the position and momentum observables; thus, 
one may think that the noncommutative-geometrical picture lurking behind 
quantum mechanics can only be interpreted as some sort of phase space, rather 
than a space(time). For a theory of particle dynamics, physical space is observable 
only as the space of all possible positions for a free particle, or (more generally) 
the center of mass for a closed system of particles, {\em i.e.} the configuration 
space. There is, however, no notion of configuration space in quantum mechanics. 
In the classical case, we have the configuration space being given by ``half'' of 
the phase space; therefore, it makes good sense to then look carefully at the 
phase space. Our analysis from the relativity symmetry perspective \cite{066}
answers this question well: we have illustrated that the Hilbert space -- as the
quantum  phase space -- can be seen as reducing to either the Newtonian phase
space \textit{or} configuration space in the classical limit, formulated as the 
symmetry contraction limit. Such a contraction is the ``inverse'' of a relativity 
symmetry deformation. To put it another way, the phase space is an irreducible 
representation under the quantum relativity symmetry, while the classical 
phase space is reducible to the configuration space and momentum space 
parts. Einsteinian special relativity says Newtonian space and time are only 
parts of Minkowski spacetime, and can be handled separately only under 
the Newtonian approximation. Our result gives a similar picture regarding 
the position and momentum parts of the phase space. Consequently, 
the phase space is the right model for physical space. Some other authors 
working on noncommutative spacetime have indeed brought in a 
symplectic or Poisson structure \cite{Y1} for noncommutative spacetime,
through not connecting the latter with the phase spaces of classical
and quantum mechanics. Our perspective here suggests thinking more
about deformation of spacetime and its dynamics through  deformation
of what used to be called phase space instead of what is essentially
only the configuration space part. It is interesting to note that the notion 
of the phase space as the inseparable model of the physical spacetime 
in quantum settings actually has a parallel in the recent considerations 
of Born reciprocity and doubled geometry from string theory \cite{born}.
It remains to see though if there may be the necessity to go beyond the 
symplectic and Riemannian geometric setting from our framework.

The projective Hilbert space, as the space of pure states, is a dual 
geometric structure to the observable algebra \cite{S}. It is still a 
commutative manifold, but an infinite-dimensional one. We consider 
it an alternative, real-number geometric description of the 
noncommutative geometry. The quantum space behind quantum 
mechanics may be described by the six $\hat{X}_i$ and $\hat{P}_i$ 
noncommutative coordinate observables or the infinite set of real 
number coordinates of the projective Hilbert space.  Both sets 
essentially reduce to the same classical limit of the Newtonian 
phase space coordinates $x_i$ and $p_i$.

The Weyl-Wigner-Groenewold-Moyal (WWGM) formalism
 \cite{W,H,GV,D,Z}, or deformation quantization, is a key approach 
for passing from the commutative to the noncommutative. Such a 
deformation quantization of coordinate observables has been a theme 
in the construction of noncommutative spacetime.  The present paper, 
however, rewrites the WWGM formalism from the relativity symmetry
perspective, i.e. from the cyclic irreducible representation of the group 
$C^*$-algebra corresponding the the unitary representation of the 
group on the Hilbert space. The classical limit is retrieved as the symmetry
contraction limit pushing all the way to the algebra of observables and their 
dynamical evolution. Deformation quantization is therefore established 
as the deformation of the representation of the group $C^*$-algebra, 
arising as a consequence of the relativity symmetry deformation. We see 
this work as providing the crucial first link from the bottom-up to any 
quantum/noncommutaive models of spacetime. It also suggests looking 
at quantum physics from a noncommutative-geometric point of view. 
Most, if not all, of the mathematics presented in the paper is essentially 
there scattered in the physics and mathematics literature. Our work
is to pull all that together under an explicit consistent framework
to illustrate our $\tilde{G}(3)$ relativity perspectives for the particle 
dynamics as one on the quantum model of the physical space, and
the full passage to $G(3) [\otimes U(1)]$ Newtonian limit. 

A word on some of the mathematics not otherwise addressed here
may be in order. Deformation quantization in the some references 
to noncommutative geometry is written in the language of the twisting 
of a Hopf algebra structure. Nothing like the latter has been used in the
case of basic quantum mechanics, which is the case considered in the 
analysis presented in this article. That said, the group $C^*$-algebra 
for the Heisenberg-Weyl group, or its irreducible representations, 
has a natural (twisted) convolution as a noncommutative product
to which the Moyal star-product is essentially a Fourier transform.
The group $C^*$-algebra can also be promoted to a Hopf algebra
by a simple commutative coproduct, as such it is dual to the Hopf
algebra with the commutative pointwise function product. It also 
possesses a natural noncommutative coproduct \cite{K}. In the
commutative contraction limit of the group, all of these products
and coproducts are, of course, commutative.

\section{ WWGM Formalism as Representation of Group C*-algebra Corresponding to Coherent State Representation}
The Weyl-Wigner-Groenewold-Moyal (WWGM) formalism \cite{W,H,GV,D,Z} 
describes quantum mechanics on the space of tempered distributions 
on the set of variables, usually taken as coordinates of the classical 
phase space. Physical observables, as hermitian operators on the Hilbert 
space of pure states, are described by real functions. A quantum state, 
pure or mixed, corresponds exactly to a function for the density operator. 
A noncommutative product - the Moyal star product - gives the space of 
generalized observables the structure of an algebra. The Moyal algebra for
physical observables is isomorphic to the corresponding algebra of operators
and reduces to the Poisson algebra of classical observables in the $\hbar\to 0$
limit. While the formalism provides a complete description of quantum
mechanics without the need for the usual wavefunctions, it is usually -- and 
especially historically -- introduced through the Weyl-Wigner transform 
starting from Schr\"odinger wavefunction over position variables. Such
wavefunctions is only background used for deriving the results, but is not
an integral part of the formulation. The (canonical) coherent state
 \cite{csr1,csr2,csr3,csr4}
representation provides an alternative giving the Schr\"odinger wavefunction
as a function on essentially `classical' phase space variables much similar to 
the operators in WWGM itself. The symplectic manifold of the set of classical 
states can be seen as the submanifold of the basis coherent states in the 
infinite-dimensional K\"ahler manifold of the quantum projective Hilbert space 
\cite{BZ}. Moreover, there is the Koopman-von Neumann formulation of 
classical mechanics in the language of Hilbert spaces \cite{KN1,KN2, cHsp}.  
From all of these, we expect looking at the WWGM formalism through 
the Weyl-Wigner transform starting instead from the Hilbert space of 
wavefunctions over the coherent state basis will provide a particularly 
interesting picture of quantum mechanics which would also be suitable 
for the analysis of the classical limit. One would then have operators
and states both plausibly described by elements in the same space of 
functions or distributions. Consequently, both the WWGM and the Hilbert 
space formalisms may be unified as one. We are not aware of any explicit 
treatment along these lines; hence we present one in this article  
(important work for the coherent states has, however, been given in 
Ref.\cite{DG}).  A key feature, for example, is that a function $\za$ acts 
by the star product $\za\star$ as an operator on a wavefunction $\phi$; 
the Moyal star product $\za\star\zb$ between functions is essentially 
the operator product $\za\!\star\zb\star$. In other words, the wavefunctions 
we start with end up as objects inside the Moyal algebra(s) of `observables'; 
the Hilbert space is nothing other than the one  obtained through an 
algebraic GNS construction \cite{Da1,Da2} from the Moyal algebra itself. 
 
In a recent paper \cite{066}, we have introduced the idea of looking at 
the central extension of the Galilean group as the true relativity symmetry 
group for quantum mechanics and the Newtonian case as the contraction 
limit of the symmetry \cite{c1,c2,060} trivializing the central extension. The 
Hilbert space serves as the phase space, and in a way also the configuration 
space, of a free quantum particle. Taken as representation spaces of the 
relativity symmetry, the limit of the two picture under the contraction 
can be traced to give what are essentially the phase and configuration space
of the classical Newtonian picture. The latter is more directly given, naively, 
in the language of Hilbert spaces; hence the Koopman-von Neumann 
formulation --- a part of the story to be described explicitly here.
The coherent states serve as the basis for the construction of the quantum 
Hilbert space representation and are the only ones surviving as (pure) states
in the classical limit. The current study is partly motivated, therefore, by the 
need for a corresponding full dynamical picture of this story. Indeed, the 
Heisenberg picture provides a more transparent illustration of the dynamics. 
The WWGM formalism is, of course, supposed to focus on the observables 
more than the pure states as vectors in a Hilbert space. The coherent state 
formulation, however, makes the pure states directly accessible within 
the same algebraic framework. This framework gives a solid setting for 
the analysis of time evolution both within and outside of the relativity 
symmetry perspective. All symmetries can be described on a similar footing, 
as unitary transformations on the Hilbert space and automorphisms on the 
algebra of observables. 

We especially want to emphasize here our perspective that all of the 
mathematical structures behind the physical theory naturally manifest 
themselves from the (relativity) symmetry group and its associated structures. 
The Hilbert space of pure states is an irreducible unitary representation of
the group and the observable algebra is an irreducible representation of the
group ($C^*$-)algebra. The former as an irreducible representation of the 
latter, within our framework, sits naturally inside the latter; the natural 
(noncommutative algebraic) multiplicative actions of which is the operator
action. For a somewhat parallel picture for describing mixed states, we bring 
in the notion of a Tomita representation \cite{TT,trep}, which sees a density 
matrix (for a mixed state) as a vector in a Hilbert space (of operators). This 
is particularly useful for describing symmetry in the Koopman-von Neumann 
formulation in the symmetry contraction limit. Symmetries are represented 
as unitary transformations on the Hilbert spaces and inner automorphisms 
of the observable algebra. The dynamical picture naturally follows. 
So, the analysis here establishes explicitly that the $\tilde{G}(3)$ plays 
the complete role of a relativity symmetry to quantum mechanics and
give the classical approximation of the Newtonian case in all 
corresponding aspects.

In light of the above, the basic perspective of our framework is to start 
with the coherent state representation, essentially seen as a representation 
of the extended Galilean symmetry, which is equivalent to the one formulated 
simply with the Heisenberg-Weyl group \cite{066}, and the classical theory 
is to be retrieved through the contraction limit. We write quantum expressions
with the $\hbar=2$ units here, following Ref.\cite{GV}. This choice gives 
the Heisenberg commutation algebra the `unusual' form of 
\bea
[X_i, P_j]=2i\delta_{ij} I \;,
\eea
while setting the minimal uncertainty product to unity. It is in many ways 
the more natural choice of units for quantum theory. The next section sets 
the stage for the Hilbert space description. The explicit WWGM formalism 
is presented in Sec.~IV. We emphasize the group theoretical aspects, and give 
more details only for results and features specific to the coherent state 
framework. Sec.~V presents the symmetry contraction to the classical theory.
The Tomita representation is introduced at the end of the section. In Sec.~VI, 
we summarize the description of quantum symmetries and work out explicit results 
for elements of the relativity symmetry, including results on the Hilbert space 
of the Tomita representation. Dynamics is described by a time (translation
symmetry) transformation. The section following this traces the symmetry 
descriptions to the contraction limit, and Sec.~VIII focuses on the classical 
dynamical picture {\em a la} Koopman-von Neumann (and otherwise). 
The last section presents some concluding remarks.

\section{The Coherent State Representation}
We start with the familiar (canonical) coherent state representation 
\bea
\left|p^i,x^i\rra = {U}(p^i,x^i) \left|0\rra \equiv
e^{-i\theta} U(p^i,x^i,\theta) \left|0\rra
\eea
where
\bea
U(p^i,x^i,\theta) \equiv 
e^{ix_i p^i \hat{I}}
e^{i\theta \hat{I}} 
e^{-ix^i\hat{P}_i} 
e^{ip^i\hat{X}_i} 
= e^{i(p^i\hat{X}_i- x^i\hat{P}_i +\theta\hat{I})} \;,
\eea
$\left|0\rra \equiv \left|0,0\rra$ is a fiducial normalized vector, 
$\hat{X}_i$ and $\hat{P}_i$ are representations of the generators $X_i$ 
and $P_i$ as self-adjoint operators on the abstract Hilbert space  $\mathcal{H}$ 
spanned by the $\left|p^i,x^i\rra$ vectors, and 
$\hat{I}$ is the identity operator representing the central generator $I$. 
 $(p^i,x^i,\theta)$ corresponds to a generic element of the 
Heisenberg-Weyl group as  
 \bea
{W}(p^i,x^i,\theta) = \exp [i(p^iX_i- x^iP_i +\theta I)],
\eea
with
\bea\label{ww}
{W}(p'^i, x'^i, \theta')  {W}(p^i, x^i, \theta)
= {W}\!\!\left(p'^i+p^i, x'^i+x^i, \theta'+\theta-(x'_i p^i - p'_i x^i) \right) \;.
\eea
Here, $(x'_i p^i - p'_i x^i) $  is the classical mechanical symplectic form.  
Note that we have $p^i$ and $x^i$ here corresponding to {\em half} the 
expectation values of $\hat{P}_i$  and $\hat{X}_i$ ($\hbar=2$ units).
In the interest of simplifying the notation, we drop the index $i$ in most
of the subsequent expressions, only putting it back explicitly when some
emphasis on the three-vector nature of a given quantity is warranted.

 We introduce wavefunctions on the coherent state
manifold $\phi (p,x )\equiv \lla p,x| \phi \rra$ with 
\bea 
 \lla p,x \left| \hat{X} \right|\phi \rra &=& \hat{X}^{\!\ssc L}  \phi (p,x ) \;,
\nonumber \\
\lla p,x \left| \hat{P} \right|\phi \rra &=& \hat{P}^{\!\ssc L}  \phi (p,x ) \;,
\eea
where
\bea
\hat{X}^{\!\ssc L} &=&  x +  i \partial_{p}\;,
\nonumber \\
\hat{P}^{\!\ssc L} &=& p -  i \partial_{x}\;,
\label{L}
\eea
and
\bea \label{u-shift}
{U}^{\ssc\! L}(p,x) \phi (p',x') \equiv
 \lla p',x' \left|{U}(p,x)  \right|\phi \rra 
= \phi (p'-p,x'-x ) e^{{i}(px'-xp')}\;.
\eea
Furthermore, we have a realization of the quantum Hilbert space as a closed 
(polarization) subspace of $L^2(\Pi)$, the space of the wavefunctions 
$\phi (p,x)$ on which ${U}^{\!\ssc L}$ acts as a (projective) 
representation of the Heisenberg-Weyl group. We can see again that 
$\hat{P}^{\!\ssc L}$ and $\hat{X}^{\!\ssc L}$ generate translations in
$x$ and $p$, respectively. The wavefunction $\phi_a (p,x)$ 
for the coherent state $\left|p_a,x_a \rra$  is given by 
\bea \label{ol}
\phi_a (p,x) \equiv  \lla p,x |p_a,x_a \rra 
= e^{i(p_ax-x_ap)} e^{-\frac{1}{2}[(p-p_a)^2+(x-x_a)^2]} \;;
\eea
in particular, the $\left| 0,0\rra$ state wavefunction is denoted by
$\phi_0 (p,x)$ and $\phi_0 (p,x)= e^{-\frac{1}{2}(p^ip_i+x^ix_i)}$, which is  
a symmetric Gaussian of unit width. The expression $ \lla p,x |p_a,x_a \rra$ 
may also be taken as giving the overlap of two different coherent states. 
$\phi_a (p,x)$ is a test function belonging to the Schwartz space of smooth 
rapidly decreasing functions $S(\Pi)$. In what follows, we will denote the 
Hilbert space  of wavefunctions by $\mathcal{K}$. The representations, 
with or without the superscript $L$ 
({\em i.e.} on  $\mathcal{K}$ or on $\mathcal{H}$), are of course 
unitarily equivalent.  The natural inner product on $\mathcal{K}$ is 
 $\frac{1}{\pi^n}  \int\!\! dp dx \bar{\phi}(p,x){\phi'}(p,x)
 =\langle {\phi}| \hat{I} |\phi'\rangle$ with
 $\hat{I}=\frac{1}{\pi^n} \int dp dx \left|p,x \rra \!\!\lla p,x \right|$,
which keeps $\phi_a (p,x)$  as a normalized wavefunction.   
Note that we use $n$ for the dimension of the classical physical
space, though we only consider $n=3$ here. 

For a full discussion of the algebra of smooth observables, we will go 
beyond the Hilbert space of pure states for the limited class of bounded
observables. Pure states for smooth observables are unit rays in
$\mathcal{S} \equiv {S}(\Pi) \!\cap\! \mathcal{K}$ \cite{D}, 
though at times we may not pay full attention to the difference 
below.

\section{The Observable Algebra from the WWGM Formalism}
In this article, we emphasize the key role of the associated structures 
of the symmetry group behind the physical theory.  We have seen that the 
Heisenberg-Weyl group manifold, or the isomorphic coset space of the 
extended Galilean group, provides a direct description of the coherent state 
basis \cite{066} for the Hilbert space. Here, we see how the group ring  
provides a description of  the set of operators and the observable algebra. 
The set of operators 
\bea
\Omega'[\za(p,x,\theta)] =\frac{1}{(2\pi)^{n+1}} 
\int \!\!dp dx d\theta \za(p,x,\theta) U(p,x,\theta) \;,
\eea
where $\za(p,x,\theta)$ is a distribution on the group manifold, is a 
Heisenberg-Weyl ring \cite{W}. Consider $\za(p,x)\equiv \za^1(p,x)$ 
in the expansion
\bea \label{za}
\za(p,x,\theta)= \int \!\!d\lambda \, |\lambda|^n \za^{\lambda} (p,x) e^{-i\lambda\theta} \;.
\eea
It can easily be seen that the $\theta$-integration in $\Omega'[\za(p,x,\theta)]$
can be performed to give a $2\pi \delta (\lambda-1)$, which in turn yields
$\Omega'[\za(p,x,\theta)] =\Omega[\za(p,x)]$ with
\bea
\Omega[\za(p,x)] =\frac{1}{(2\pi)^n}
\int \!\!dp dx  \,\za(p,x) {U}(p,x) \;.
\eea
$\lambda$ can be interpreted as an eigenvalue of the central charge 
generator which is always unity under the representation.
We have the latter as a continuous linear injection from $L^1(\Pi)$ into 
${I\!\!B}(\mathcal{H})$. As such, it is a *-algebra homomorphism with 
respect to the twisted convolution product $\circ$ and the involution 
$*$ defined by
\bea
\za \circ \zb (p,x) =\frac{1}{(2\pi)^n}  \int\!\! dp' dx'  \,\za (p',x')  \zb(p -p',x-x')\,
e^{{i}(p'x-x'p)}
\eea
and 
\bea
\za^*  = \overline{\za(-p,-x)} \;,
\eea
respectively, where $\bar{\za}$ is the complex conjugate of $\za$.
That is, $\Omega[\za \circ \zb ]=\Omega[\za] \Omega[\zb ]$ 
and  $\Omega[{\za}^*]=\Omega[\za]^\dag$. 
Self-adjoint elements of ${I\!\!B}(\mathcal{H})$ and their counterparts 
in $L^1(\Pi)$ represent the bounded observables. Here, $\Pi$ is the 
$(p,x)$-space, which can be considered as the coherent state 
manifold and is also the `classical phase space' ${I\!\!R}^{2n}$ on 
which we have the wavefunctions $\phi(p,x)$. Note that the $\Omega$ 
map takes a delta function $\delta_a$ of mass 1 centered on 
$(p_a, x_a)$ to ${U}(p_a, x_a)=e^{i(p_a\hat{X}- x_a\hat{P})}$,
and $\za \circ \delta_o = \delta_o\circ \za =\za$ for the delta 
function $\delta_o$ centered on $(0,0)$.
The inverse mapping can be written as 
\bea
\za(p,x)= {2^n}\mbox{Tr} \big[\Omega[\za] {U}^\dag(p,x)\big] \;,
\eea
where the trace is to be evaluated over the set of
coherent states as 
$ \frac{1}{\pi^n}\int\!\! dp''dx'' \lla p'',x''| \cdot |p'',x''\rra$
and we have 
\bea \label{tr}
\mbox{Tr}  [{U}(p',x') {U}^\dag(p,x)]
   ={\pi^n} \delta(p'-p,x'-x)  \lla p,x | p',x' \rra \;.
\eea

Similarly, we have 
\bea
\Omega^{\ssc L}[\za(p',x')] =\frac{1}{(2\pi)^n}
\int \!\!dp'dx'  \,\za(p',x') {U}^{\ssc L}(p',x') \;,
\eea
where ${U}^{\!\ssc L}(p',x')=e^{i(p'\hat{X}^{\!\ssc L}- x'\hat{P}^{\!\ssc L})}
=e^{i(p'x-x'p)} \, e^{-(p'\partial_p + x'\partial_x)}$,
with the set of $\Omega^{\ssc L}[\za]$ considered 
as operators on $L^2(\Pi)$ satisfying
$\Omega^{\ssc L}[\za \circ \zb ]=\Omega^{\ssc L}[\za] \Omega^{\ssc L}[\zb ]$.
Naturally
\footnote{Here
$\mbox{Tr} [\hat{\za}^{({\ssc L})}]=\frac{1}{\pi^n} \int\!\! dp_adx_a \,
 \bar{\phi_a} \hat{\za}^{({\ssc L})} {\phi_a}$.
}
\bea
\za(p,x)= {2^n} \mbox{Tr} \big[\Omega^{\ssc L}[\za] {U}^{{\ssc \!L}\dag}(p,x)\big] \;.
\eea
In fact, the left-invariant vector fields, or differential operator 
realization of the generators, of the group manifold 
\bea
X^{\!\ssc L} &=&i  x  \partial_\theta +  i \partial_{p}\;,
\nonumber \\
P^{\!\ssc L} &=& i p  \partial_\theta -  i \partial_{x}\;,
\nonumber \\
I^{\!\ssc L} &=& i \partial_\theta \;,
\eea
have their action on a function $\za(p,x,\theta)$ in the form of Eq.(\ref{za}) 
given by the action on $\za^{\lambda} (p,x)$ defined by ${\lambda}x+i \partial_{p}$, 
${\lambda}p-i \partial_{p}$, and $0$, respectively. Hence, ${\lambda}=1$ 
yields the differential operators $\hat{X}^{\!\ssc L}$ and  $\hat{P}^{\!\ssc L}$ 
of Eq.(\ref{L}), as in $U^{{\ssc \!L}\dag}(p,x)$, which are exactly the 
corresponding operators acting on $\za(p,x) [\equiv \za^1(p,x) ]$.

The symplectic  Fourier transform
\bea
\za_f(p,x)= F[\za](p,x) \equiv \frac{1}{(2\pi)^n} 
  \int\!\! dp'dx' \,\za(p',x') \, e^{i(p'x -x'p)} \;.
\eea
is a continuous isomorphism of $S(\Pi)$, as a Fr\'echet space, onto itself
extending to a unitary transformation on $L^2(\Pi)$ with $F^2=1$.
The twisted product $\star$ satisfies
\bea
F[\za] \star F[\zb] 
= F[\za \circ \zb]  \;
\eea 
and
\footnote{We use $F^{-1}$ instead of simply $F$ to keep track of 
difference which only manifests at the classical contraction limit
discussed in the next section.}
\bea
F^{-1}[\za \star \zb] =  F^{-1}[\za] \circ F^{-1}[\zb]\;,
\eea 
with $1 \star \za=\za=  \delta_o \circ \za$, in which case the two 
products commute. We also have
\bea \label{c2s}
\za\circ\zb = F[\za]\star\zb \;.
\eea
This is the usual Moyal star product, which can be written as
\bea
\za \star \zb (p,x) = \za(p,x) e^{-i (\cev{\partial}_p \vec{\partial}_x-\cev{\partial}_x \vec{\partial}_p) } \zb(p,x) \;,
\eea
or in the integral form
\bea
\za \star \zb (p,x) = \frac{1}{(2\pi)^{2n}} \int\!\! dp'dx' dp''dx''  \za(p',x')  \zb(p'',x'') 
                          e^{ -i(px'-xp')}e^{ i(px''-xp'')}e^{ i(p''x'-x''p')} .
\eea
In particular, we have 
\bea
x \star \za &=& (x+i\partial_p) \za = \hat{X}^{\!\ssc L} \za  \;,
\nonumber \\
p \star \za &=& (p-i\partial_x) \za = \hat{P}^{\!\ssc L} \za \;.
\label{xpstar}
\eea
The Fourier transform $F$ is a continuous *-algebra isomorphism between 
$[S(\Pi), \circ, \dag]$ and $[S(\Pi), \star, \bar{}~]$, the latter involution 
being simple complex conjugation. Both the $\star$ and $\circ$ products can be 
extended to the space $S'(\Pi)$ of tempered distributions. Notice that $F$ is 
more commonly written as a transform between functions of two different spaces, 
one being the parameter space for the Heisenberg-Weyl group modulo $\theta$, 
while the other is identified as the classical phase space, or rather the 
variable space of the Moyal star functional algebra. Our perspective of looking 
at quantum mechanics and its classical limit, focusing on the coherent state 
picture \cite{066}, may be considered a justification for identifying the
two, as done in Refs.\cite{H,GV} for example, at the quantum level. 
The `classical phase space' then is also the coherent state manifold with 
parameters characterizing, however, half the position and momentum 
expectation values
\footnote{Note that though it looks like we have inconveniently made the 
group parameters and the coherent state expectation values differ by a factor 
of $2$ by using $\hbar=2$ instead of $\hbar=1$ units, it is really results like
Eq.(\ref{xpstar})  that naturally prefer the convention. The parameter space
for the wavefunctions $\phi$ can be exactly identified with that of the 
Moyal star functional algebra.
}.

Consider
\bea
\Delta^{\!\!({\ssc L})} [\za] \equiv \Omega^{({\ssc L})} [F^{-1}[\za]] &=& \frac{1}{(2\pi)^{2n}}
\int \!\! dp' dx'dp dx  \,\za(p,x) e^{i(px'-xp')} {U}^{({\ssc L})}\!(p',x') 
\nonumber \\
&=& \frac{1}{(2\pi)^n} \int \!\!dp dx  \,\za(p,x) \Delta_{p,x}^{\!\!({\ssc L})}\;,
\eea
where we have 
\footnote{We have
$\Delta_{p,x}^{\!\!({\ssc L})}  = {U}^{({\ssc L})}\!(p,x)  \Delta_{0,0}^{\!\!({\ssc L})}$ 
with $\Delta_{0,0}^{\!\!({\ssc L})}$ being the phase space parity operator of
Grossmann-Royer \cite{GR1,GR2}; {\em i.e.} $\Delta_{0,0}\left|p',x'\rra=\left|-p',-x'\rra$
and $\Delta_{0,0}^{\!\!{\ssc L}} \phi(p',x') =\phi(-p',-x')$.
Note that $\Delta_{p,x}^{\!\!({\ssc L})}$ is actually selfadjoint, besides being unitary.
}
\bea
 \Delta_{p,x}^{\!\!({\ssc L})} =\frac{1}{(2\pi)^n}   \int\!\! dp'dx' \, e^{i(px'-xp')} {U}^{({\ssc L})}\!(p',x') \;.
\eea
In these expressions, we are putting the two cases, with and without the superscript 
$L$, in a single set. This is  the  Weyl correspondence, {\em i.e.} we have 
$\hat\za^{\ssc L}\! \equiv\za(\hat{P}^{\!\ssc L},\hat{X}^{\!\ssc L})=\Delta^{\!\ssc L}[\za(p,x)] $
and $\hat\za\! \equiv\za(\hat{P},\hat{X})=\Delta[\za(p,x)] $ with, 
however, $\hat\za^{\ssc L}$ here thought of as operators on wavefunctions on the 
manifold of coherent states. Then the bijective map $\Delta^{\!{\ssc L}}$ 
takes $S'(\Pi)$ to $\mathcal{L}(S(\Pi),S'(\Pi))$, and we have
\bea
\za(p,x)= 2^n \mbox{Tr} [\hat{\za}^{({\ssc L})}\Delta_{p,x}^{\!\!({\ssc L})\dag}] \;.
\eea
Moreover, $\Delta^{\!({\ssc L})}[\za \star \zb (p,x) ]
     =\Delta^{\!({\ssc L})}[\za(p,x) ] \Delta^{\!({\ssc L})}[\zb(p,x) ]$
gives a *-algebra isomorphism between the Moyal algebra 
$[\mathcal{M}, \star, \bar{}\, ]$ and the corresponding algebra of 
smooth `observables' ${\mathcal L}^+(S(\Pi))$, with 
$\mathcal{M} \equiv\{\zb \in  S'(\Pi) : \zb \star \za \,,
    \za \star \zb \in S(\Pi) \;\forall \za \in S(\Pi)\}$; 
and between $[\mathcal{M}', \star, \bar{}\, ]$ and ${\mathcal L}^+(S(\Pi),L^2(\Pi))$
as algebra of bounded `observables', with
$\mathcal{M}':= \{\zb \in  S'(\Pi) : \zb \star \za \,,
    \za \star \zb \in L^2(\Pi) \;\forall \za \in L^2(\Pi)\}$.
Note that $\Delta[\bar\za(p,x) ]=\Delta[\za(p,x) ]^\dag$; hence physical 
observables with $\hat{\za}=\hat{\za}^\dag$, are given by real elements
of the Moyal algebras. We will, however, mostly not pay much attention to 
the difference between $\mathcal{M}$ and $\mathcal{M}'$ below. 

We define ($\star$-)multiplicative operators acting on the distributions by
\bea \label{M}
M_\star[\za] \equiv \za \star   \;.
\eea
Then, we have the simple and elegant result
\bea
M_\star[\za(p,x)] = \Delta^{\!\ssc L}[\za(p,x)] 
= \hat{\za}^{{\ssc L}}=\za(\hat{P}^{\!\ssc L}, \hat{X}^{\!\ssc L})
\eea
which can be interpreted as the Bopp shift.
The representation given through $\hat{X}^{\ssc L}$ and $\hat{P}^{\ssc L}$ of 
Eq.(\ref{L}) on $\mathcal{K}$ directly extends to arbitrary functions $\za(p,x)$  
and coincides with the $\star$ product structure  with $\hat{\za} \left|\phi\rra$
described in $\mathcal{K}$ as $\hat{\za}^{\ssc L} \phi= \za \!\star \phi$ and
$\za\!\star\zb\star=(\za\!\star\zb)\star$ as 
$\za\!\star (\zb\!\star \phi)=(\za\!\star\zb)\!\star \phi$. It is the left 
regular representation of the functional algebra on itself, which can be 
extended further to all of $S'(\Pi)$. One can even associate the wavefunction 
$\phi$ with a $\hat\phi^{\ssc L}= \phi\,\star$ operator, though the latter 
does not correspond to a physical observable. It remains to be seen if the 
operator has any particular physical meaning. Looking at a real wavefunction, 
or $|\phi|$, makes more sense as the absolute phase of a quantum state 
has no physical meaning anyway. $|\phi|\star$ makes a legitimate 
physical observable. We are interested only in applying all these mathematical 
results to the Gelfand triple $\mathcal{S} < \mathcal{K} < \mathcal{S}'$
and that is the background on which the explicit results concerning the
states are to be understood. We will see at the end that $\mathcal{K}$
is essentially the left ideal of $L^2(\Pi)$ that carries an irreducible 
representation of the Moyal algebra.

The Wigner functions that describe states, pure or mixed, are to be given in 
terms of  functions $\rho$ of the density operator $\hat\rho$. For a pure state 
$\left|\phi\rra$, the latter is denoted by 
$\hat\rho_\phi \equiv  \left|\phi\rra\!\!\lla \phi\right|$. Two different pure
states give $\hat\rho_{\phi\phi'} \equiv  \left|\phi'\rra\!\!\lla \phi\right|$, 
with a non-diagonal Wigner function given as
$\rho_{\phi\phi'}=2^n \mbox{Tr} [\hat\rho_{\phi\phi'} \Delta_{p,x}^\dag]$.
Focusing on the set of basis coherent states, one can check that actually
$F[\phi_a]=\phi_a$ and $F[\bar\phi_a]=\bar\phi_{-a}$.
For $\hat\rho_{ab} \equiv\left|p_b,x_b\rra\!\!\lla p_a,x_a \right|$,
we have, with Eq.(\ref{c2s}), 
\bea \label{rab}
\rho_{ab}(p,x) =
2^{2n} \phi_b \circ \bar\phi_a =2^{2n} F[\phi_b ]\star \bar\phi_a 
      =2^{2n} \phi_b \star \bar\phi_a \;.
\eea
Explicitly,
\bea \label{rabe}
\rho_{ab}(p,x)
= 2^n e^{i(p_ax_b-x_ap_b)} e^{i(p_bx-x_bp)} e^{-i(p_ax-x_ap)}
e^{-\frac{(p-p_a-p_b)^2+(x-x_a-x_b)^2}{2}} \;,
\eea
which for $b=a$ reduces to 
\bea 
\rho_{a}(p,x) =  2^n e^{-\frac{(p-2p_a)^2+(x-2x_a)^2}{2} }
 \;.
\eea
The latter is the Wigner function for the coherent state $\left|p_a,x_a\rra$,
a Gaussian of unit width centered at $(2p_a,2x_a)$ the point with coordinates 
exactly at the expectation values. We can then obtain
\bea
\mbox{Tr}  [ \hat\za^{\!{\ssc L}} ] &=&\frac{1}{\pi^n}  \int\!\! dp_adx_a \, 
  \frac{1}{\pi^n}  \int\!\! dpdx \, \bar\phi_a (\za\star \phi_a)
\\
&=& \frac{1}{2^n(2\pi)^n}  \int\!\! dpdx \, \za
 \frac{1}{\pi^n} \int\!\! dp_adx_a \,   \rho_a
=\frac{1}{2^n(2\pi)^n}   \int\!\! dpdx \, \za \;,
\eea
in which we have used the associativity and trace properties of the star product.
This result corresponds to the standard trace expression  for $\za$, or rather
$\za\star$ ($\za=\za\star 1$). We also denote this by $\mbox{Tr}  [\za]$ for
simplicity. From a transition amplitude, we have
\bea
\frac{1}{\pi^n}   \int\!\! dpdx \,  \za ( \phi' \star \bar\phi)
= \mbox{Tr}  [ \za\star \rho_{\phi\phi'}\star] 
=\frac{1}{2^n} \frac{1}{(2\pi)^n}   \int\!\! dpdx \,  \za \rho_{\phi\phi'} \;.
\eea
The latter result includes as special cases the standard 
$\lla\phi | \hat{\za} | \phi \rra= \mbox{Tr} [\za \rho_{\phi} ]$
and the somewhat strange looking
$\frac{1}{\pi^n}   \int\!\! dpdx  \bar\phi \phi
=\frac{1}{2^n}\frac{1}{(2\pi)^n}   \int\!\! dpdx   \rho_{\phi}$.
\footnote{
It is interesting to see the consistency of this result for the explicit case of a 
coherent state $\phi_a$. The normalization condition for a wavefunction in 
$\mathcal{K}$ can be written in the form
\bea
1&=&\frac{1}{\pi^n}  \int\!\! dp dx \bar{\phi}\phi
=\frac{1}{2^n} \frac{1}{(2\pi)^n}  
    \int\!\! d({2}p) d({2}x) \, e^{-\frac{1}{2}[2(p-p_a)^2+2(x-x_a)^2]}
\nonumber \\ \nonumber 
&=& \frac{1}{2^n}\frac{1}{(2\pi)^n}  \int\!\! d\tilde{p}^s d\tilde{x}^s \,
   e^{-\frac{1}{8}[2(\tilde{p}^s-2p_a)^2+2(\tilde{x}^s-2x_a)^2]} \;,
\eea
to be compared with
\bea\nonumber 
1=\frac{1}{2^n} \frac{1}{(2\pi)^n}   \int\!\! dpdx   \rho_a=
 \frac{1}{(2\pi)^n}  \int\!\! dp dx \, e^{-\frac{1}{2}[(p-2p_a)^2+(x-2x_a)^2]} \;.
\eea
In terms of the new variables we have
$\phi_a(\tilde{p}^s,\tilde{x}^s)=e^{\frac{i}{2}(p_a\tilde{x}^s- x_a\tilde{p}^s)} 
   e^{-\frac{1}{8}[(\tilde{p}^s-2p_a)^2+(\tilde{x}^s-2x_a)^2]}$,
a Gaussian centered at the expectation values $(2p_a,2x_a)$ with width $\frac{1}{2}$.
}
Actually, we have 
\bea \label{w2d}
\rho_{\phi\phi'}= 2^{2n}\phi' \circ \bar\phi 
     =2^{2n}F[\phi']\star \bar\phi =2^{2n}\phi'\star \bar\phi\;,
\eea
which may  be considered as following from Eq.(\ref{rab}), since any state
(wavefunction) can be taken as a linear combination of the $\phi_a$ basis 
states. One can further check explicitly that 
$\mbox{Tr} [\rho_{a}^2]=
\mbox{Tr}  [\rho_{a}] =1$,
$\rho_{a} \star \rho_{a} =\rho_{a}$, and $\rho_{a} \star \phi_{a} =\phi_{a}$
for the functions $\rho_{a}$ and $\phi_{a}$. Another interesting result is
\bea
[\za\!\star\phi](p_a,x_a)= \frac{1}{\pi^n} 
   \int\!\! dp_b dx_b \, \mbox{Tr}  [\za \rho_{ab}]  \phi(p_b,x_b) \;,
\eea
which is the key result in Ref.\cite{DG} giving, together with Eq.(\ref{rabe}), 
the explicit integral kernel of the operator $\za\star$; more specifically 
we have
\bea
\phi(p_a,x_a)= \frac{1}{\pi^n}  \int\!\! dp_b dx_b \, \mbox{Tr}  [\rho_{ab}]  \phi(p_b,x_b) \;.
\eea
The set of $\rho_{ab}$ spans the Hilbert space $L^2(\Pi)$, or equivalently the set of
$\rho_{ab}\star$ spans  ${\mathcal{T}}_2(\mathcal{K})$, which is the space of
Hilbert-Schmidt operators, with the inner product
$\lla\!\lla \za|\zb\rra\!\rra=\mbox{Tr}[\bar\za\zb]$.
$[ L^2(\Pi), \star, ^-]$ is a generalized Hilbert algebra \cite{TT} and
$[ S(\Pi), \star, ^-]$ a subalgebra.

In the algebraic formulation on $S(\Pi) < L^2(\Pi)< S'(\Pi)$, the (normalized)
states are to be defined by the positive and normalized functionals on 
${\mathcal L}^+(S(\Pi))$ given by $\omega_{\!\rho} (\za)= \mbox{Tr}[\za \rho]$; 
hence, essentially the set of density operator functions $\rho$ (we will use 
the term density matrix for such a function below). Each $\rho_\phi\star$ is 
a projection operator, and the one-dimensional projections correspond to 
pure states $\omega_\phi\equiv\omega_{\!\rho_{\!\phi}}$. The Hilbert 
space is to be constructed as the completion of the quotient 
${\mathcal I}_{\mathcal K}/{\mathcal I}_o$ of closed left ideals  
with a (pre-)inner product on ${\mathcal I}_{\mathcal K}$ through an 
$\omega_\phi$ to which ${\mathcal I}_o$ is the kernel \cite{D,Da1,Da2}.
An obvious choice here is the Hilbert space
${\mathcal I}_{\mathcal K}=\{\za\!\star\rho_o : \za\in L^2(\Pi) \}$ and
${\mathcal I}_o$ of $[ L^2(\Pi), \star, ^-]$ with the inherited inner product 
$\lla \za | \zb \rra_{\!\omega_o}
  \equiv \lla\!\lla \za\!\star\rho_o | \zb\!\star\rho_o \rra\!\rra
   =\omega_o(\bar{\za}\!\star \!\zb)$ for which ${\mathcal I}_o=\{0\}$. 
Note that the inner product is exactly the same as
$\langle \hat\za\phi_o | \hat\zb\phi_o \rangle$ illustrating that
the Hilbert space is equivalent to ${\mathcal H}$ or ${\mathcal K}$.
 In fact, $\rho_o=2^n\phi_o$; hence ${\mathcal K}$ and
${\mathcal I}_{\mathcal K}$ differ only in normalizations.
${\mathcal I}_{\mathcal K}$ is invariant under the action of the Moyal
algebra ${\mathcal M}'$ and as a representation it is a faithful 
and irreducible one, which matches with it being irreducible as a 
representation of the Heisenberg-Weyl symmetry. After all, the algebra 
of observables is to include the enveloping algebra of the latter. 
Any particular wavefunction $\phi$ of ${\mathcal K}$ can be used
to give a representation through $\omega_\phi$ which are all 
unitary equivalent. So, we come full circle to the representation
on ${\mathcal K}$ giving an explicit illustration of the more abstract
algebraic language through the WWGM framework.

\section{Lie Algebra Contraction Limit}
Consider the contraction \cite{c1,c2} of the Lie algebra for the  Heisenberg-Weyl
subgroup of the full relativity symmetry given by the $k \to \infty$ limit with
 the rescaled generators 
\bea
X_i^c = \frac{\sqrt{\hbar}}{k} X_i 
\qquad \mbox{and} \qquad
\label{XPc}
P_i^c =  \frac{\sqrt{\hbar}}{k} P_i \;. 
\eea
We have $[X_i^c, P_j^c]= i \frac{2\hbar}{k^2}\delta_{ij} I \to 0 (\hbar)$. Here, 
$k$ is a pure numerical parameter while $\hbar$ is Planck's constant, which is 
needed to allow $X_i^c$ and $P_j^c$ to take on independent physical units, 
such as the usual classical units. One can take the parameter $k$ here as $\sqrt{2}$ 
to give the standard form of the commutator $[X_i^c, P_j^c]= i\hbar I$ and think 
of the contraction limit as the $\hbar \to 0$ limit, i.e. as the classical limit is 
usually envisaged. However, as discussed in Ref.\cite{066},  naively taking 
$\hbar$ to zero everywhere in the theory written with $\hbar$ carrying 
nontrivial physical units is not the right thing to do; case examples can be 
seen below. On the other hand, the contraction limit is of course independent 
of the contraction parameter used, and the physical units for $X_i^c$ and 
$P_i^c$ differ from those of  $X_i$ and  $P_i$ by the same factor of the physical
unit (that of $\sqrt\hbar$), even in the limit. To keep track of things carefully, 
in a way that enables the reader to see expressions in their familiar forms 
with a nonzero $\hbar$ as well as the contraction limit results, we keep 
$\hbar$ and $k$ separate in this section. As we mentioned above, substituting 
$\sqrt{2}$ for $k$ yields the familiar quantum expressions with their proper 
$\hbar$ dependence, and then we can interpret the naive choice of taking 
$\hbar\to 0$ (which can be interpreted in the classical system of units) as the 
classical limit; however, we instead take $k \to \infty$ as the appropriate 
choice for describing the classical limit in the symmetry (representation) 
contraction perspective. 

We first rewrite the Heisenberg-Weyl group element in the usual form
${W}(\breve{p}^i,\breve{x}^i,\theta) 
  = e^{\frac{i}{\hbar}(\breve{p}^i X_i^c- \breve{x}^i P_i^c +\theta \hbar I)}$,
where
\bea
\breve{p}^i = \sqrt\hbar k \, p^i\;,
\qquad \mbox{and} \qquad 
\breve{x}^i = \sqrt\hbar k \, x^i\;.
\eea
Here, $\hbar$ is introduced so that $\breve{p}^i$ and $\breve{x}^i$ 
carry the same units as $P_i^c$ and $X_i^c$, respectively. 
As for the invariant vector fields acting on the group manifold, we have
\bea
{X^c}^{\!\ssc L} =   \frac{\breve{x}}{k^2}    i \partial_\theta 
  +   i \hbar \partial_{\breve{p}}  &\quad\longrightarrow&  i \hbar \partial_{\breve{p}}  \;,
\nonumber \\
{P^c}^{\!\ssc L} =  \frac{\breve{p}}{k^2}     i \partial_\theta 
  -   i \hbar \partial_{\breve{x}} &\quad\longrightarrow& - i \hbar \partial_{\breve{x}} \;.
\label{go-c}
\eea
Note that the contraction limits of these vector fields, as given above, carry 
a factor of $\hbar$. This is needed in order to have the correct physical units; 
the dropping out of the terms  $i \partial_\theta$ is to be expected, as the 
central charge $I$ fully decouples from the rest of the algebra. In particular, 
observe that, had we replaced $k^2$ by 2, taking the naive $\hbar \to 0$ 
limit would yield an incorrect result with \textit{only} $i \partial_\theta$ 
in the classical limit. 

To trace the contraction of the coherent state representation, simply 
relabeling the set of $\left|p,x\rra$ states  by 
$\left|\breve{p},\breve{x}\rra= {U}(\breve{p},\breve{x})  \left| 0 \rra$
 with ${U}(\breve{p},\breve{x}) 
     = e^{\frac{i}{\hbar}(\breve{p}^i \hat{X}_i^c- \breve{x}^i \hat{P}_i^c)}$
is {\em not} the right thing to do. It should be emphasized that the basic
idea for taking the classical approximation as the symmetry contraction limit 
is to take the original representation which describes the quantum physics
to the required limit rather than directly building the classical physics
description from the contracted symmetry. We will see that the contraction of
the representation does indeed give a representation of the contracted symmetry
though. The essence of the coherent state representation is to have the states 
labeled essentially by their (finite) expectation values. However, we have
$x_i^s \equiv \lla \breve{p},\breve{x} \big|\hat{X}_i^c \big| \breve{p},\breve{x}  \rra
    =   \frac{2}{k^2} \breve{x}_i$   
and similarly  $p_i^s= \frac{2}{k^2} \breve{p}_i$. Let us instead take 
$\left|{p}^s,{x}^s\rra$ and note that
\bea
p^s = \frac{2\sqrt\hbar}{k} \, p \;,
\qquad \mbox{and} \qquad 
x^s= \frac{2\sqrt\hbar}{k} \, x \;.
\eea
On the Hilbert space $\mathcal{K}$ of wavefunctions ,  we then have
\bea
\hat{X}^{c\ssc L} =    \frac{x^s}{2}+   i \frac{2\hbar}{k^2} \partial_{p^s} &\quad\longrightarrow&  \frac{x^s}{2}\;,
\nonumber \\
\hat{P}^{c\ssc L} =   \frac{p^s}{2}  -  i \frac{2\hbar}{k^2} \partial_{x^s} &\quad\longrightarrow&  \frac{p^s}{2}\;.
\eea
The difference between these results and those of Eq.(\ref{go-c}) is remarkable. 
The overlap between the different coherent states 
$\lla {p}^{s\prime},{x}^{s\prime} |{p}^s,{x}^s\rra$ can be obtained easily from 
Eq.(\ref{ol}). As discussed in Ref.\cite{066}, it vanishes at the contraction limit 
and $\mathcal{H}$, as a representation for the observable algebra, which is to 
be interpreted  as functions of $\hat{X}^c$ and  $\hat{P}^c$ reducing to a 
simple sum of the one-dimensional spaces for each $\left|{p}^s,{x}^s\rra$. On 
$\mathcal{K}$, the coherent state wavefunctions $\phi_a({p}^s,{x}^s)$ are each, 
apart from a phase factor, a Gaussian with width $\frac{\sqrt{2\hbar}}{k}$, and 
hence collapse to the delta function $\delta_a({p}^s,{x}^s)$ in the limit. The 
classical coset space picture \cite{066} of the Newtonian phase space can be 
considered as having been retrieved. Sticking to the Hilbert space picture,  
$\phi_a({p}^s,{x}^s) \to \delta_a({p}^s,{x}^s)$ makes $\mathcal{K}$ the whole 
of $L^2(\Pi)\equiv L^2(\Pi^s)$, which is the Koopman Hilbert space for classical 
mechanics \cite{cHsp}.  A word of caution is warranted on the integration 
measure. The normalization condition of $\phi_0({p}^s,{x}^s)$ reads
\bea
 \left( \frac{k^2}{4 \pi\hbar} \right)^{\!\!n} 
     \int\!\!  d{p}^s d{x}^s \; e^{-\frac{k^2}{4\hbar} [(p^s)^2+(x^s)^2]} =1 \;,
\eea
the measure of which apparently diverges at the $k\to\infty$ limit. Similarly,
$\hat{I}= \frac{1}{\pi^n} dpdx \left|{p},{x}\rra \!\!\lla {p},{x} \right|
=(\frac{k^2}{4\hbar \pi})^n \int\!\!  d{p}^s d{x}^s \left|{p}^s,{x}^s\rra \!\!\lla {p}^s,{x}^s \right|$.
The delta function limit has the measure $\frac{1}{(2\pi\hbar)^n} \int\!\!  d{p}^s d{x}^s$
instead. As the $k$-dependence drops at the $k\to \infty$ limit, we need
the $\hbar$ to fix the physical units. The set of $\phi_a({p}^s,{x}^s)$, then
as $\delta_a$, as a basis for $\mathcal{K}$ is maintained. However, a
generic function as a linear combination of $\phi_a({p}^s,{x}^s)$ loses the
status of being a physical state at least as far as pure states are concerned. 
The basis coherent states, or rather their classical limit as given by
$\delta_a({p}^s,{x}^s)$ (or simply points $({p}^s,{x}^s)$ of the familiar
classical phase spaces), are the only pure states.

Next, we look at all of the above integral transformations in order to track 
the Weyl correspondence at the contraction limit. Here, it is not so clear 
whether we should use the group parameters $\breve{p}$ and $\breve{x}$ 
or the coherent state parameters. In fact, both work. Note that in the 
discussion of the coherent state above, in view of things in the last 
section, it is better to use $\left|{p}^c,{x}^c\rra$ with 
\bea
p^c = \frac{p^s}{2} =\frac{\sqrt\hbar}{k} \, p \;,
\qquad \mbox{and} \qquad 
x^c = \frac{x^s}{2} =\frac{\sqrt\hbar}{k} \, x \;
\eea
instead, especially as that gives the ratio between the old and new parameters 
as the same as that between the operators [{\em cf.} Eq.(\ref{XPc})]. $p^s$
and $x^s$ were mostly used there for easy reading of familiar results 
(at $k^2=2$). Explicitly, we have
\bea
\hat{X}^{c\ssc L} &=&    x^c +   i \frac{1}{k^2} \partial_{p^c} \longrightarrow x^c\;,
\nonumber \\
\hat{P}^{c\ssc L} &=&   p^c -   i \frac{1}{k^2} \partial_{x^c}\longrightarrow p^c\;,
\eea
giving a nice contraction limit. That is of course nothing more than a 
convenient choice of convention,  one we adopted in the 
beginning through labeling the coherent states by half their
expectation values. In terms of the new parameters, we have
\bea
\Omega^{({\ssc L})}[\za] =\left(\frac{k^2}{2\pi\hbar}\right)^{\!\!n}  \!\!
\int \!\! d{p}^c d{x}^c  \,\za({p}^c,{x}^c) {U}^{({\ssc L})}({p}^c, {x}^c) \;,
\eea
and
\bea
F[\za] =\left(\frac{k^2}{2\pi\hbar}\right)^{\!\!n}  \!\!
\int \!\! d{p}^c d{x}^c  \,\za({p}^c,{x}^c) e^{\frac{ik^2}{\hbar}({p}^c x^{c\prime} - {x}^c p^{c\prime})}  \;.
\eea
Note that -- despite the factor of $k^{2n}$  in front -- the latter two expressions 
are $k$-independent. The factor of $k^{2n}$ cancels after integration; note also that
${U}^{({\ssc L})}({p}^c, {x}^c)=e^{\frac{ik^2}{\hbar}[{p}^c \hat{X}^{(\ssc L)} - {x}^c \hat{P}^{(\ssc L)}]}$. 
 It follows that
\bea
\Delta^{\!({\ssc L})} [\za] = \Omega^{({\ssc L})} [ F^{-1}[\za]] 
&=& \left(\frac{k^2}{2\pi\hbar}\right)^{\!\!2n} \!\!
   \int \!\!dp^c dx^c dp^{c\prime} dx^{c\prime} \,\za(p^c,x^c) 
e^{\frac{ik^2}{\hbar}(x^{c\prime}p^c-p^{c\prime}x^c)} 
   {U}^{({\ssc L})}\!(p^{c\prime},x^{c\prime}) 
\nonumber \\
&=& \left(\frac{k^2}{2\pi\hbar}\right)^{\!\!n}  \!\!
  \int \!\!dp^c dx^c  \,\za(p^c,x^c) \Delta_{p^c,x^c}^{\!({\ssc L})}\;,
\eea
 where we have 
\bea
 \Delta_{p,x}^{\!({\ssc L})} &=& \left(\frac{k^2}{2\pi\hbar}\right)^{\!\!n}  \!\!
  \int \!\! dp^{c\prime} dx^{c\prime} \, e^{\frac{ik^2}{\hbar}(x^{c\prime}p^c-p^{c\prime}x^c)} 
   {U}^{({\ssc L})}\!(p^{c\prime},x^{c\prime})  \;.
\eea
The Weyl correspondence is obviously maintained\footnote{
One may also consider
\bea\nonumber 
\Omega^{c({\ssc L})}[\za] \equiv \frac{1}{(2\pi\hbar)^n}
\int \!\! d\breve{p} d\breve{x}  \,\za(\breve{p},\breve{x}) {U}^{({\ssc L})}(\breve{p}, \breve{x}) \;,
\eea
and
\bea\nonumber 
F^c  [\za] \equiv \frac{1}{(2\pi\hbar)^n} 
  \int\!\! d\breve{p} d\breve{x} \,\za(\breve{p},\breve{x}) \, 
  e^{\frac{i}{\hbar}(\breve{p} x^c - \breve{x} p^c)} \;.
\eea
Actually, we have $\Omega^{c({\ssc L})}[\za]=k^{2n}  \, \Omega^{({\ssc L})}[\za]$
and $F^c  [\za]= k^{2n} \, F[\za]$ formally ($F^{c-1}\ne F^c$). It follows that
\bea
\Delta^{\!({\ssc L})} [\za] = \Omega^{c({\ssc L})} [ F^{c-1}[\za]] 
&=& \frac{1}{(2\pi\hbar)^{2n}}
   \int \!\!dp^c dx^c d\breve{p} d\breve{x}  \,\za(p^c,x^c) e^{\frac{i}{\hbar}(\breve{x}p^c-\breve{p}x^c)} 
   {U}^{({\ssc L})}\!(\breve{p},\breve{x}) 
\nonumber \\ \nonumber 
&=& \frac{1}{(2\pi\hbar)^n} \int \!\!dp^c dx^c  \,\za(p^c,x^c) \Delta_{p^c,x^c}^{\!c({\ssc L})}\;,
\eea
with
$\Delta_{p^c,x^c}^{\!c({\ssc L})}
= \frac{1}{(2\pi\hbar)^n}   \int\!\! d\breve{p} d\breve{x} \,
    e^{\frac{i}{\hbar}(\breve{x}p - \breve{p}x )} {U}^{({\ssc L})}\!(\breve{p},\breve{x}) \;
 [ = k^{2n} \,\Delta_{p,x}^{\!({\ssc L})}] $
The twisted convolution  required to maintain
$\Omega^{c({\ssc L})}[\za \circ^c \zb ]=\Omega^{c({\ssc L})}[\za] \Omega^{c({\ssc L})}[\zb ]$
is simply given formally by $k^{2n} \za \circ \zb$, {\em i.e.}
\bea \nonumber 
\za \circ^c \zb (\breve{p},\breve{x}) =\frac{1}{(2\pi\hbar)^n} \int\!\! d\breve{p}' d\breve{x}'  \,
\za (\breve{p}',\breve{x}')  \zb(\breve{p} -\breve{p}',\breve{x}-\breve{x}')\,
e^{\frac{i}{k^2\hbar}(\breve{p}\breve{x}'-\breve{x}\breve{p}')}  \;.
\eea
};
explicitly, we have
$\Delta [\za(p^c,x^c)]= \za(\hat{P}^{c}, \hat{X}^{c})$ and 
$\Delta^{\!{\ssc L}}\! [\za(p^c,x^c)]= \za(\hat{P}^{c{\ssc L}}, \hat{X}^{c{\ssc L}})$.
Analogous trace formulas for the inversion of the transforms work as well.
With the original $\circ$ with $\Omega^{({\ssc L})}$ written in terms of the 
parameters $p^c$ and $x^c$, we have the Moyal star product given by
\bea \label{starc}
\star^c \sim
\exp \bigg[ \frac{-i\hbar}{k^2} (\cev{\partial}_{p^c}  \vec{\partial}_{x^c}  
    -\cev{\partial}_{x^c} \vec{\partial}_{p^c} ) \bigg],
\eea
which is the same old expression written in terms of the new parameters
serving as arguments for the (observable) functions, though we denote it by
$\star^c$ for convenience of later referral. The star product reduces to 
the simple product in the $k\to\infty$ limit, as it should. The equation 
$M_\star[\za(p^c,x^c)] 
=\za(p^c,x^c)\star^c$
is maintained for all finite $k$ and at the $k\to\infty$ limit where the
$M_\star$ becomes the simple multiplication of the then classical observable
$\za(p^c,x^c)$. The suggestive notation of the multiplicative operator
$M[\za(p^c,x^c)]$, or simply $M_\za$, is standard for an operator on the 
$L^2(\Pi)$ Hilbert space representing classical observables in the 
Koopman-von Neumann formulation of classical mechanics \cite{cHsp}.

The coherent state Wigner function is
\bea 
\rho_{a}(p^c,x^c) = 2^n e^{-\frac{k^2}{\hbar}\frac{(p^c-2p_a^c)^2-(x^c-2x_a^c)^2}{2} } \; ,
\eea
with width $\frac{\sqrt{\hbar}}{k}$, and it reduces to  a delta function 
as $k\to\infty$. For the classical case, any `density matrix' $\rho(p,x)$ 
beyond the delta functions are to be interpreted as statistical
distributions; hence they are mixed states. 
A generic $\rho(p,x)$ has the form
\bea
\rho(p,x)=\sum_m c_m \rho_{\phi_m} (p.x)
=2^{2n}\sum_m c_m [\phi_m \star \bar\phi_m] (p.x)
\eea
for pure states $\phi_m (p.x) \in \mathcal{K}$, and $c_m$ are positive
real numbers with $\sum_m c_m =1$.  Therefore, we have
\bea
\rho(p^c,x^c) = 2^{2n}\sum_m c_m [\phi_m \star^c\bar\phi_m] (p^c,x^c)
\to 2^{2n}\sum_m c_m |\phi_m (p^c,x^c) |^2 
\eea
as $k\to\infty$. To be more rigorous, one would have to introduce a
rescaled/renormalized $\rho^c$ for the limiting classical density matrix
as the conventional distribution to describe the classical statistical state
through a bounded $\rho^c(p^c,x^c)= \mbox{lim}_{k\to \infty} k^{2n} \rho(p^c,x^c)$
in order to maintain $\frac{1}{(2\pi\hbar)^n} \!\int \!\!\rho^c dp^cdx^c =1$.
The classical wavefunction $\phi^c$, satisfying $|\phi^c|^2\equiv\rho^c$, 
for the Koopman-von Neumann formulation may then be introduced. It is
important to emphasize that a  function $\phi^c$ describes a mixed state.
Only the limiting distributions of the delta functions describes the
(classical) pure states.  

An even better formal picture of the classical limit, which is also particularly 
useful for the description of dynamics and symmetries below, is offered by
the notion of a Tomita representation \cite{TT} as presented in Ref.\cite{trep}.
Note that the representation is not an irreducible one -- an aspect that fits
the Koopman-von Neumann formulation well. We present only a specific 
description based on the wavefunctions $\phi(p,x)$, although also using
the $\left|\phi\rra$ notation when it is more illustrative and convenient.
Consider the Hilbert space $\tilde{\mathcal{K}}$ of square ket vectors 
$\left|{\za}\right]$ defined as the span of all 
$\left|\phi'\phi\right] \equiv\rho_{\phi\phi'}$ with the inner product
given by
\bea
\left[\psi'\psi|\phi'\phi\right] = \mbox{Tr} [ \bar\rho_{\psi\psi'}\rho_{\phi\phi'}]\;.
\eea
$\tilde{\mathcal{K}}$ is thus a tensor product of ${\mathcal{K}}$ with itself.
It is essentially the space of Hilbert-Schmidt operators on ${\mathcal{K}}$,
and can be thought of as the span of all $\rho_{ab}$. 
A vector $\left|{\za}\right]$ corresponds to $\za\in L^2(\Pi)$ or the
Hilbert-Schmidt operator $\za\star$ on ${\mathcal{K}}$ and we can identify it
with $\left.\left|{\za}\rra\!\rra$, introduced towards the end of the last section.
That is, within our formulation,   ${\mathcal{K}}$ can be identified as  $L^2(\Pi)$.
Next, one can introduce the conjugation $J$ as an antiunitary operator with the properties:
\bea (i)&&
J c \left|{\za}\right] = \bar{c} J  \left|{\za}\right] 
\quad \forall c \in\mathds{C} \;,
\nonumber \\ (ii)&&
\left[{\za}|J^\dag J |{\zc}\right] = \left[{\gamma} \left|\right.{\za}\right]  \;,
\nonumber \\ (iii)&&
J^2=I \;.
\eea
State vectors in $\tilde{\mathcal{K}}$ are introduced as vectors in a self-dual 
cone $\tilde{\mathcal{D}}$ of  with real and positive inner products. The vectors
correspond to the mixed states. They satisfy 
$J\!\left|{\rho}\right]=\left|{\rho}\right]\equiv \rho$. 
For any operator $\za\star$ on ${\mathcal{K}}$, we have an operator
$(\za \star)_{\!\ssc T}$ on $\tilde{\mathcal{K}}$ defined by
\bea
(\za \star)_{\!\ssc T} \left |\phi'\phi\right] 
\equiv \left |(\za\!\star\!\phi')\phi\right]=\rho_{\phi (\za\star\phi')}
 =\za\!\star \rho_{\phi\phi'}\;,
\eea
as $\rho_{\phi\phi'} =2^{2n}\phi'\star\bar{\phi}$. 
We have $J\left |\phi'\phi\right] =\left |\phi\phi'\right]$, hence
 $J (\za \star)_{\!\ssc T}  J \rho_{\phi\phi'}=\rho_{(\za\star\phi)\phi'}$.
Note that $(\za \star)_{\!\ssc T}$ and $J (\za \star)_{\!\ssc T}  J$
act on different parts of the tensor product; explicitly
$(\za\star)_{\!\ssc T}=(\za\star) \otimes 1$ and
$J (\za \star)_{\!\ssc T}  J = 1 \otimes (\za\star)$. More explicitly,
we have
\bea
J \zb =\bar\zb \;,
\eea
and
\bea
(\za \star)_{\!\ssc T}  \zb & =& \za\!\star \zb\;,
\nonumber \\ \label{LR}
J (\za \star)_{\!\ssc T} J \zb &=& \zb \star \bar{\za} \;,
\eea
where we have extended the results for $\rho_{\phi\phi'}$ to a generic
$\zb \in  \tilde{\mathcal{K}}$, and utilized some algebraic properties of the star
product, including $\overline{\za\star\zb}=\bar{\zb} \star \bar{\za}$. 
A pure state is then represented by 
$\left |{\rho}_\phi\right] \equiv\left |\phi\phi\right]$ 
and we have $\left[{\rho}_\psi|{\rho}_\phi\right]=  \mbox{Tr} [ \rho_{\psi}\rho_{\phi}]
     =|\!\lla\psi|\phi\rra\!|^2$; hence $\left |{\rho}_\phi\right]$
is normalized. The expectation value 
$\left[{\rho}_\phi \left|(\za \star)_{\!\ssc T} \right|{\rho}_\phi\right]$ 
for a pure state is the same as on ${\mathcal{K}}$, {\em i.e.} equal to
$\lla\phi |\hat{\za}|\phi\rra$ and $\mbox{Tr}[\za\rho_\phi]$; 
and for a mixed state 
$\left[ {\rho} \left| (\za \star)_{\!\ssc T} \right|{\rho}\right] = \mbox{Tr}[\za\rho]$.
Transition probabilities are given by
$\left[{\rho}_\psi \left|(\za \star)_{\!\ssc T}  J (\za \star)_{\!\ssc T}  J\right|
   {\rho}_\phi\right]= |\!\lla \psi|\hat{\za}\phi\rra\!|^2=
    \left(\mbox{Tr}[\za\rho_{\psi\phi}]\right)^2$.   
Note that $\left|{\rho}\right] \in \tilde{\mathcal{D}}$ always gives
$\left.\left|(\za \star)_{\!\ssc T}  J (\za \star)_{\!\ssc T}  J\right|
   {\rho}\right] \in  \tilde{\mathcal{D}}$. Observables on $\tilde{\mathcal{D}}$
are to be taken from those within the $(\za\star)_{\!\ssc T}$ and
$J (\za \star)_{\!\ssc T} J $ set. Other vectors outside $\tilde{\mathcal{D}}$ 
and operators on $\tilde{\mathcal{K}}$ beyond this collection are not of much interest.

As $\tilde{\mathcal{K}}$  is essentially the space of Hilbert-Schmidt operators
on ${\mathcal{K}}$, the classical picture from the contraction limit of the latter
as a representation of the Heisenberg-Weyl symmetry obviously maintains
the basic notion of the Hilbert space as $L^2(\Pi)$ to be coordinated by the
classical phase space variables $(p^c,x^c)$, though a renormalization may
again be necessary  to trace them from the explicit original quantum 
$\za(\star)$. The original $\tilde{\mathcal{D}}$ is really the real span of 
$\rho_a$ for the coherent state basis; hence it becomes the real span of the
delta functions in the classical limit. That is to say, the set of classical `density 
matrices' fills the whole real part of $L^2(\Pi)$. Formally, $\tilde{\mathcal{D}}^c$ 
is simply the real part of $\tilde{\mathcal{K}}^c$.  We can see further that
$(\za \star)_{\!\ssc T} \to M_\za$ and $ J (\za \star)_{\!\ssc T}  J \to M_{\bar\za}$.
More features of the classical picture obtained will be seen in Sec. VII below,
where we discuss the description of dynamics.

\section{Description of Quantum Symmetries and Time Evolution}
The description of the quantum symmetries in connection with the WWGM 
formalism has been well presented in Ref.\cite{BCW1,BCW2}, from which we 
summarize the basic features and give explicit details for applications to our 
framework with particular emphasis on the elements of the relativity symmetry. 
Hermitian operators, as physical observables, play the role of the symmetry 
generators giving rise to a group of unitary flow on the Hilbert space(s), as well 
as an isomorphic group of automorphisms on the set of pure state density 
operators. Here we only focus on ${\mathcal{K}}$ and the matching set of 
$\rho_\phi$, while extending from the latter to all of
$\tilde{\mathcal{D}} \in\tilde{\mathcal{K}}$ in the language of the
Tomita representation.

Firstly, we have on ${\mathcal{K}}$ symmetries as the group of unitary and 
antiunitary operators factored by its closed center of phase transformations. 
The isomorphic automorphism group $Aut({\mathcal{P}})$ of the set 
${\mathcal{P}}$ of $\rho_\phi$ is characterized by the subgroup of the 
group of real unitary transformations ${\mathcal{O}}(\tilde{\mathcal{K}}_{\!\ssc R})$
compatible with the star product, {\em i.e.} $\mu \in Aut({\mathcal{P}})$
satisfies
\bea
\mu(\za\star\zb) =\mu(\za)\star\mu(\zb) \;
\eea
[or $\mu(\za\!\star\zb\star) =\mu(\za)\!\!\star \mu(\zb)\star $]. 
$\tilde{\mathcal{K}}_{\!\ssc R}$ is the real subspace of the Hilbert space of
Hilbert-Schmidt operators.   Symmetry groups represented as subgroups of 
$Aut({\mathcal{P}})$ can be considered. For a star-unitary transformation 
$U_\star$ on  wavefunctions $\phi \in {\mathcal{K}}$,  we have a real 
unitary operator on $\tilde{\mathcal{K}}_{\!\ssc R}$
\bea \label{ug}
 \tilde{U}_\star\za\star=\mu(\za)\star=U_\star\! \star \za  \star  \bar{U}_\star \;.
\eea

We write a generic one parameter group of such a (star-)unitary transformation
in terms of real parameter $s$ as ${U}_{\star}\!(s)\star=e^{\frac{-i s}{2}{G}_{\!s}\star}$
with ${G}_{\!s}\star$ as the generator. Note that the factor $2$ is put in the place 
of $\hbar$, consistent with our choice of units
\footnote{For readers who find the factor of $2$ difficult to appreciate,
our results below in the next section -- especially with the symmetry description in terms of the
rescaled parameters in the usual units with an explicit $\hbar$ -- should make the full picture more transparent.}.
For time translation, as a unitary transformation on ${\mathcal{K}}$,
we have the Schr\"odinger equation of motion 
\bea
2i \frac{d}{dt} \phi = {G}_{\!t}\!\star \phi \;.
\eea
In the Tomita representation picture,  the unitary transformation 
${U}_{\star}(s)\star$ on ${\mathcal{K}}$ gives a corresponding unitary 
transformation on $\tilde{\mathcal{K}}$ as the Hilbert space of square kets, 
which are here simply elements in $L^2(\Pi)$, given by \cite{trep}
\bea
\tilde{U}_{\star}\!(s)= {U}_{\star}(s) J {U}_{\star}(s) J
=e^{\frac{-i s}{2}\big[({G}_{\!s}\star)_{\!\ssc T} - J ({G}_{\!s}\star)_{\!\ssc T}  J \big]} \;.
\eea
From Eq.(\ref{LR}), one can see that this is just a fancy restatement of 
Eq.(\ref{ug}) above, with now an explicit form of $\tilde{U}_{\star}\!(s)$ 
as an operator in terms of the real generator function ${G}_{\!s}(p,x)$.
Consider the generator $\tilde{G}_{\!s}$ as defined by 
$\tilde{U}_{\star}\!(s)= e^{\frac{-i s}{2}\tilde{G}_{\!s}}$, we have 
\bea
\tilde{G}_{\!s} \rho \equiv
 \tilde{G}_{\!s} \left |{\rho}\right]
    =\big[({G}_{\!s}\star)_{\!\ssc T} - J ({G}_{\!s}\star)_{\!\ssc T}  J \big] \left |{\rho}\right]
    ={G}_{\!s}\!\star \rho -\rho \star\!{G}_{\!s} 
  =\{{G}_{\!s},  \rho\}_\star\;,
\eea 
where $\{\cdot, \cdot\}_\star $ is the star product commutator, {\em i.e.} the 
Moyal bracket. Hence, with $\rho(s)=\tilde{U}_{\star}\!(s) \rho(s=0)$, 
\bea \label{leq}
\frac{d}{ds}\rho(s)=\frac{1}{2i} \{{G}_{\!s},  \rho(s)\}_\star \;.
\eea
The result is of course to be expected. When applied to the time translation as 
a unitary transformation, it gives exactly the Liouville equation of motion for 
a density matrix as the Schr\"odinger equation for the latter taken as a state in  
$\tilde{\mathcal{D}}$ with Hamiltonian operator $\tilde{\kappa}=\tilde{G}_{t}$.
$G_{\!s}(p^i,x^i)\star=G_{\!s}(p_i\star,x_i\star)=G_{\!s}(\hat{P}_i^{\!\ssc L},\hat{X}_i^{\!\ssc L})$ 
is the operator that represents the (real) element $G_{\!s}({P}_i,{X}_i)$ in  the 
algebra of observables as well as $\mathcal{K}$. 
$\tilde{G}_{\!s}=\{{G}_{\!s},  \cdot\}_\star$ then represents the algebra 
element on $\tilde{\mathcal{K}}$.  It is interesting to note that by introducing 
the notation  $\hat{G}_{\!s}^{\ssc L}\equiv G_{\!s}(p^i,x^i)\star$ as a left 
action, we have a corresponding right action $\hat{G}_{\!s}^{\ssc R}$
given by $\hat{G}_{\!s}^{\ssc R}\za=\za\star\!{G}_{\!s}$ with
$\tilde{G}_{\!s}=\hat{G}_{\!s}^{\ssc L}-\hat{G}_{\!s}^{\ssc R}$.
The explicit expression for the $\hat{G}_{\!s}^{\ssc R}$ action follows from
\bea
\hat{X}^{\!\ssc R}_i &=&  x_i -  i \partial_{p^i}\;,
\nonumber \\
\hat{P}^{\!\ssc R}_i &=& p_i +  i \partial_{x^i}\;.
\label{R}
\eea
These operators match to the right-invariant vector fields of the 
Heisenberg-Weyl group as $\hat{X}^{\!\ssc L}$ and $\hat{P}^{\!\ssc L}$
corresponds to the left-invariant ones.
When $G_{\!s}(p,x)$ is an order one or two polynomial in the variables, 
which covers the cases of interest here, $\tilde{G}_{\!s}$ has a
very simple explicit form. Another important feature to note is that
$\tilde{G}_{\!s}$ determines  $G_{\!s}(p,x)$ only up to an additive constant.
This is a consequence of the fact that the density matrix $\rho_\phi$ is
insensitive to the phase of the pure state $\phi$. Constant functions in the
observable algebra correspond to multiples of $\hat{I}$ on  $\mathcal{H}$ 
which generates pure phase transformations, 
{\em i.e.} $G_{\!\theta}(p,x)=1$ and $\tilde{G}_{\!s}=0$. 

Let us focus first on the observables $x\star$ and $p\star$ as symmetry 
generators on $\mathcal{K}$, and look at the corresponding transformations 
on $\tilde{\mathcal{D}}$. From Eq.(\ref{u-shift}), 
\bea
U_{\star}\!(x)\!\star \phi (p',x')& =& e^{\frac{-i x}{2} (-p\star)} \phi (p',x')= \phi \!\left(p',x'+\frac{x}{2} \right) e^{{i}xp'}\;,
\nonumber \\
U_{\star}\!(p)\!\star \phi (p',x')& =& e^{\frac{-i p}{2}(x\star)} \phi (p',x') =  \phi \!\left(p'+\frac{p}{2},x' \right) e^{{-i}px'}\;,
\eea
giving, in terms of explicit $x^i$ and $p^i$ parameters, 
\bea
G_{\!-x^i}\star &=& p_i\star\;, \qquad\qquad \tilde{p}_i = \tilde{G}_{\!-x^i} =2i \partial_{x^i} \;,
\nonumber \\
G_{\!p^i}\star &=& x_i\star \;, \qquad\qquad   \tilde{x}_i =\tilde{G}_{\!p^i} = 2i \partial_{p^i} \;.
\eea
The factors of $2$ in the translations $U_{\star}\!(x)\!\star$ and 
$U_{\star}\!(p)\!\star$ may look somewhat suspicious at first sight. They 
are actually related to the fact that the arguments of the wavefunction 
correspond to half the expectation values, due to our coherent state labeling. 
Thus, we find that $x\star$ and $p\star$ generate translations of the 
expectation values, which is certainly the right feature to have. To better appreciate 
these results, one can also think about a sort of `Heisenberg picture' for the 
symmetry transformations as translations of the observables instead of 
states giving the same transformations of the expectation values.
One can see that the differential operators play an important role as 
operators on $\tilde{\mathcal{K}}$. We can consider the set of $x_i$, $p_i$
$\tilde{x}_i$ and  $\tilde{p}_i$ as the fundamental set of operators -- functions of 
which essentially describe the full algebra of observables -- versus the case on 
${\mathcal{K}}$ for which the set is given only by $x_i\star$ and $p_i\star$. 
Note that the only nonzero commutators among the set are given by
\bea
[x_i,\tilde{p}_j]=[p_i,\tilde{x}_j ]= -2i\delta_{ij} \;.
\eea
The similar fundamental set of operators was long ago introduced within 
the Koopman-von Neumann formulation \cite{L}. We see here the analogous
structure in the quantum setting. For a generic $\za(p^i,x^i)$, the function
itself ({\em i.e.} the simple multiplicative action $M_\za$), $\hat\za^{\!\ssc L}$, 
$\hat\za^{\!\ssc R}$, and $\tilde{\za}$ are all operators to be considered on 
$\tilde{\mathcal{K}}$, though only two among the four are linearly independent.

Consider $G_{\!\omega^{ij}}=(x_ip_j-x_j p_i)$. We have
\bea
G_{\!\omega^{ij}}\star &=& (x_ip_j -i x_i \partial_{x^j} + i p_j \partial_{p^i}
    + \partial_{x^j} \partial_{p^i}) - (i \leftrightarrow j)\;,
\nonumber \\
\tilde{G}_{\!\omega^{ij}}  &=& -2i (x_i \partial_{x^j}- p_j \partial_{p^i} )- (i \leftrightarrow j)\;.
\eea
with the explicit action 
\bea
U_{\star}\!(\omega^{ij})\!\star \phi (p',x')& =& e^{\frac{-i\omega^{ij}}{2} (G_{\!\omega^{ij}}\star)} \phi (p',x')
   = \lla p',x' \left|e^{\frac{-i\omega^{ij}}{2}  \widehat{G}_{\!\omega^{ij}} } \right|\phi\rra
    = \phi \!\left(\! e^{\frac{i\omega^{ij}}{2}  \widehat{G}_{\!\omega^{ij}} }[p',x'] \!\right) ,
\eea
where $\widehat{G}_{\!\omega^{ij}} =\hat{X}_i \hat{P}_j -\hat{X}_j \hat{P}_i$ is
the angular momentum operator on the Hilbert space ${\mathcal{H}}$ and 
$e^{\frac{i\omega^{ij}}{2} \widehat{G}_{\!\omega^{ij}} }[p',x']$ (no sum over the $i,j$
indices) the rotated $(p'x')$. This corresponds to the coset space action \cite{066},
 {\em i.e.} a rotation about the $k$-th direction of both the $p$ and $x$ as 
three dimensional vectors
\footnote{Note that $\widehat{G}_{\!\omega^{ij}}$ carries the units of $\hbar$,
which are taken as $2$. Hence, for the dimensionless parameter $\omega^{ij}$,
${\frac{i\omega^{ij}}{2} \widehat{G}_{\!\omega^{ij}} }$ with $2$ standing in for
$\hbar$, is the right dimensionless rotation operator.  A rotation on the $p$ or
$x$ vector corresponds to the same rotation on $2p$ or $2x$ as the expectation
values.}.
Together with the $G_{\!x^i}$ and $G_{\!-p^i}$ (and $G_\theta$) parts, we 
have the full set of operators from the generators of the $H_{\!\ssc R}(3)$ 
subgroup of the $U(1)$ extended Galilean symmetry with the time translation taken out. 
This portion of the ten generators $G_{\!s}$,  is of course a commutative set. The set 
of $G_{\!s}^{\ssc L}=G_{\!s}\star$ represents the symmetry on ${\mathcal{K}}$,
and constitute a subalgebra of the algebra of physical observables. 
We can easily see that the $G_{\!s}^{\ssc R}$ set does the same as a right action.
A $G_{\!s}^{\ssc L}$ always commutes with a $G_{\!s'}^{\ssc R}$ since, in general,
$[ \hat\za^{\!\ssc L}, \hat\zc^{\!\ssc R}]=0$. Explicitly, we have
\bea &&
[G_{\!\omega^{ij}}^{\ssc L\!/\!R}, G_{\!\omega^{hk}}^{\ssc L\!/\!R}]
   = \pm 2i( \delta_{jk} G_{\!\omega^{ih}}^{\ssc L\!/\!R}  -  \delta_{jh} G_{\!\omega^{ik}}^{\ssc L\!/\!R} 
                   +  \delta_{ih} G_{\!\omega^{jk}}^{\ssc L\!/\!R} -   \delta_{ik} G_{\!\omega^{jh}}^{\ssc L\!/\!R} ) \;,
\nonumber \\ &&
[G_{\!\omega^{ij}}^{\ssc L\!/\!R}, G_{\!-x^k}^{\ssc L\!/\!R}] = 
  \pm 2i( \delta_{jk} G_{\!-x^i}^{\ssc L\!/\!R} - \delta_{ik} G_{\!-x^j}^{\ssc L\!/\!R} ) \;,
\nonumber \\ &&
[G_{\!\omega^{ij}}^{\ssc L\!/\!R}, G_{\!p^k}^{\ssc L\!/\!R}] =
    \pm 2i( \delta_{jk} G_{\!p^i}^{\ssc L\!/\!R} - \delta_{ik} G_{\!p^j}^{\ssc L\!/\!R} ) \;,
\nonumber \\ &&
[G_{\!p^i}^{\ssc L\!/\!R}, G_{\!-x^j}^{\ssc L\!/\!R}] = \pm 2i\delta_{ij} G_\theta^{\ssc L\!/\!R} 
\nonumber \\ &&
[G_{\!p^i}^{\ssc L\!/\!R}, G_{\!p^j}^{\ssc L\!/\!R}] =
[G_{\!-x^i}^{\ssc L\!/\!R}, G_{\!-x^j}^{\ssc L\!/\!R}] = 0 \;,
\eea
and $G_\theta^{\ssc L\!/\!R}=1$ commutes with all other generators. Note that the factors of $2$
are really taking the place of $\hbar$ because of the choice of units. The upper 
and lower signs correspond to the $G_{\!s}^{\ssc L}$ and $G_{\!s}^{\ssc R}$ results,
respectively. For the $\tilde{G}_{\!s}$ set, we can see that the set of commutators 
is same as that of $G_{\!s}^{\ssc L}$ with however the vanishing $\tilde{G}_{\!\theta}$  
giving a vanishing $[\tilde{G}_{\!p^i}, \tilde{G}_{\!-x^j}]$.  As a result, we can also 
see the $\tilde{G}_{\!s}$ set without $\tilde{G}_{\!\theta}$  as giving the symmetry
without the central extension, similar to the classical case. Besides, we have
\bea &&
[{G}_{\!\omega^{ij}}, \tilde{G}_{\!\omega^{hk}}]
   = 2i( \delta_{jk} G_{\!\omega^{ih}} -  \delta_{jh} G_{\!\omega^{ik}}
                   +  \delta_{ih} G_{\!\omega^{jk}}-   \delta_{ik} G_{\!\omega^{jh}} ) \;,
\nonumber \\ &&
[{G}_{\!\omega^{ij}}, \tilde{G}_{\!-x^k}] =  -2i( \delta_{jk} G_{\!-x^i} - \delta_{ik} G_{\!-x^j} ) \;,
\nonumber \\ &&
[{G}_{\!\omega^{ij}}, \tilde{G}_{\!p^k}] = -2i( \delta_{jk} G_{\!p^i} - \delta_{ik} G_{\!p^j} ) \;,
\nonumber \\ &&
[\tilde{G}_{\!\omega^{ij}}, {G}_{\!-x^k}] = -2i( \delta_{jk} G_{\!-x^i} - \delta_{ik} G_{\!-x^j} )  \;,
\nonumber \\ &&
[\tilde{G}_{\!\omega^{ij}}, {G}_{\!p^k}] = -2i( \delta_{jk} G_{\!p^i} - \delta_{ik} G_{\!p^j} ) \;,
\nonumber \\ &&
[{G}_{\!p^i}, \tilde{G}_{\!-x^j}] =- [{G}_{\!-x^i}, \tilde{G}_{\!p^j}] = 2i \delta_{ij} \;,
\nonumber \\ &&
[{G}_{\!p^i}, \tilde{G}_{\!p^j}] =
[{G}_{\!-x^i}, \tilde{G}_{\!-x^j}] = 0\;.
\label{GtG}
\eea

The time translation generator $G_t^{\ssc L\!/\!R}$, needed to complete the above
set of ten $G_{\!s}^{\ssc L\!/\!R}$ into the full extended Galilean symmetry, 
is given by $G_{t} = \frac{p^i p_i}{2m}$ where $m$ is the particle mass.
One can see that $G_t^{\ssc L\!/\!R}$ commutes with each generator, except for having
\footnote{The usual presentation of the symmetry uses the Galilean
boost generators $K_i$ in place of $X_i$, which corresponds to
$G_{\!p^i}^{\ssc L\!/\!R}$ here. We have $K_i$ be matched to 
$ m G_{\!p^i}^{\ssc L\!/\!R}$. }
\bea
[ {G}_{\!p^i}^{\ssc L\!/\!R}, G_t^{\ssc L\!/\!R}] = \pm \frac{2i}{m} {G}_{\!-x^i}^{\ssc L\!/\!R} \;.
\eea
Similarly, we have $\tilde{G}_t=\frac{-2i}{m}  p^i \partial_{\!x^i}$ giving
$[\tilde{G}_{\!p^i}, \tilde{G}_{t}]= \frac{2i}{m}\tilde{G}_{\!-x^i}$. A generic 
Hamiltonian for a particle would have $G_{t}= \kappa$ to be given with an 
extra additive part as the potential energy $\upsilon(p,x)=\upsilon(x)$. It is also 
of some interest to illustrate explicitly the Heisenberg equation of motion in 
considerations of evolution, both in  ${\mathcal{K}}$ and $\tilde{\mathcal{K}}$.  
For the time dependent operator  $\za (p,x;t)\star$, on ${\mathcal{K}}$
we have $\frac{d}{dt}  \za\star=\frac{1}{2i} [\za \star,\kappa\star]$,
while for $\za (p,x;t)$ on $\tilde{\mathcal{K}}$ we have
\bea 
\frac{d}{dt} \za\star= \frac{1}{2i}[\za\star,\tilde{\kappa}]
    =\frac{1}{2i}(\hat\za^{\ssc L}\hat{\kappa}^{\ssc L} -\hat\za^{\ssc L}\hat{\kappa}^{\ssc R}
                           -\hat{\kappa}^{\ssc L}\hat\za^{\ssc L} +\hat{\kappa}^{\ssc R}\hat\za^{\ssc L} )
     =\frac{1}{2i} [\za \star,\kappa\star] \;,
\eea
where $\tilde{\kappa}\equiv\hat{\kappa}^{\ssc L}-\hat{\kappa}^{\ssc R}=\tilde{G}_t$;
hence, we arrive at the same equation as that on ${\mathcal{K}}$. This equation can
simply be written as
\bea \label{heK}
\frac{d}{dt}  \za =\frac{1}{2i}\{\za ,\kappa\}_\star  \;.
\eea
Taking $\tilde{\kappa}=\frac{-2i}{m}  p^i \partial_{\!x^i}+ \tilde{\upsilon}$ 
explicitly and applying this to the observables $x^i$ and $p^i$, we 
have
\bea
\frac{d}{dt} \za  = \frac{p^i}{m} \partial_{\!x^i} \za 
   - \sum_{n\; \mbox{\tiny odd}} \frac{i^{n-1}}{n !} 
(\partial_{\!p_i}^n \za)\partial_{\!x^i}^n \upsilon \;.
\eea
$\upsilon$ with vanishing third derivatives, or $\za$ with 
$\partial_{\!p_i}^n\za=0$, provide particularly important examples of
the equation $\za$ being $x^i$ and $p^i$. The equation reduces to a
form exactly the same as the one for $\za$, and $\kappa$, as
if it is a classical observable.

\section{To The Relativity Symmetry at the Classical Limit}
We have presented the formulation of the classical limit of quantum mechanics from 
the perspective of a contraction of the relativity symmetry, and the corresponding
representations, in Ref.\cite{066} within the Hilbert space picture on ${\mathcal{H}}$
and ${\mathcal{K}}$. In Sec.II above, we have presented a formulation within the
WWGM setting, focusing on the key part of the Heisenberg-Weyl subgroup. 
We are now going to push that to the full relativity symmetry.
Taking the full extended Galilean symmetry with abstract generators
$X_i$, $P_i$, $J_{ij}$, $H$ and $I$ as represented on  ${\mathcal{K}}$
by the set of eleven $G_{\!s}\!\star (=G_{\!s}^{\!\ssc L})$ above, the 
contraction is to be given by $X^c_i=\frac{\sqrt\hbar}{k} X_i$,
$P^c_i=\frac{\sqrt\hbar}{k}  P_i$, $J_{ij}^c=\frac{\hbar}{2}J_{ij}$,
$H^c=\frac{\hbar}{2} H$, and $I^c=I$ taken to the $k\to \infty$ limit. Note that 
setting $k=1$ gives the usual commutator set with an explicit $\hbar$ (in the 
place of the factor 2), which can be considered as having the generators 
in the usual, classical system of units. Again, we take the contraction of
the representation(s), with $x^c$ and $p^c$ standing in for $x$ and $p$.
As the whole algebra of $\za(p,x)\star$ reduces to the Poisson algebra
$\za(p^c,x^c)$ of classical observables $\za(p^c,x^c)$, all of the $G_{\!s}(p,x)\star$ 
yield the $G_{\!s}(p^c,x^c)$, which all commute among themselves. The
noncommutative observable algebra for the Hilbert space of pure states 
${\mathcal{K}}$, upon the symmetry contraction, reduces to a
commutative algebra as a result of the reduction of ${\mathcal{K}}$
to the simple sum of one-dimensional subspaces of each coherent state
\cite{066}. Each $\za(p^c,x^c)$ is diagonal on the resulting Hilbert space
of pure states, which contains only the delta functions. How do we recover 
the noncommutative relativity symmetry at the classical level then, either on the 
observable algebra or in the Koopman-von Neumann formulation? The answer 
is to be found from the Tomita representation picture of the Hilbert space 
$\tilde{\mathcal{K}}$. The  Koopman Hilbert space 
essentially comes from $\tilde{\mathcal{K}}$.

Take the set of $\tilde{G}_{\!s}$, we have
\bea 
\tilde{G}_{\!\omega^{ij}}^c &=& \frac{\hbar}{2} \tilde{G}_{\!\omega^{ij}}
=-i\hbar(x_i^c \partial_{x^{jc}} -p_j^c \partial_{p^{ic}} -x_j^c \partial_{x^{ic}}+p_i^c \partial_{p^{jc}}) \;,
\nonumber \\ 
\tilde{G}_{\!t}^c &=& \frac{\hbar}{2} \tilde{G}_{\!t} = \frac{-i\hbar}{m}p_i\partial_{x_i}
= \frac{-i\hbar}{m}p_i^c\partial_{x_i^c} \;,
\eea
and again $\tilde{G}_{\!\theta}^c =\tilde{G}_{\!\theta}=0$. These results are
independent of the contraction parameter $k$; in fact, they are independent of 
$\frac{p}{p^c}=\frac{x}{x^c}$. The unitary operators can
be written as
\[
\tilde{U}_{\star}(\omega)=e^{\frac{-i}{\hbar}\omega^{ij}\tilde{G}_{\!\omega}^c} \;,
\quad\quad
\tilde{U}_{\star}\!(t)=e^{\frac{-i}{\hbar}t\tilde{G}_{\!t}^c} \;.
\]
Similarly, if we take
$\tilde{G}_{\!p}^c = \frac{\sqrt\hbar}{k}\tilde{G}_{\!p}
= \frac{2i\hbar}{k^2} \partial_{p^{c}}$ and 
$\tilde{G}_{\!-x}^c = \frac{\sqrt\hbar}{k} \tilde{G}_{\!-x} 
= \frac{2i\hbar}{k^2} \partial_{x^{c}} $
(we drop the spatial index in $x$ and $p$ for simplicity, similarly for $\omega$ 
above), the results vanish in the $k \to\infty$ limit. This seems to 
create a problem, however, we are {\em not interested in} the operators 
generating translations in $p$ and $x$. We should be looking at translations 
in $p^c$ and $x^c$, {\em i.e.} rewriting $\tilde{U}_{\star}(p)$ and  
$\tilde{U}_{\star}(x)$ as $\tilde{U}_{\star}(p^c)$ and  $\tilde{U}_{\star}(x^c)$.
 Introducing generators $\tilde{G}_{\!p^c}^c$ and $\tilde{G}_{\!-x^c}^c$ satisfying
\[
e^{\frac{-i}{\hbar}p^c\tilde{G}_{\!p}^c} =\tilde{U}_{\star}(p^c) = e^{\frac{-ip}{2}\tilde{G}_{\!p}} \;,
\quad\quad
e^{\frac{-i}{\hbar}(-x^c)\tilde{G}_{\!-x}^c}=\tilde{U}_{\star}(x^c)= e^{\frac{ix}{2}\tilde{G}_{\!-x}}\;,
\]
we can see that
\bea
\tilde{G}_{\!p^c} = i\hbar \partial_{p^c} \;,
\qquad\qquad
\tilde{G}_{\!-x^c} = i\hbar \partial_{x^c} \;,
\eea
which are again independent of $\frac{p}{p^c}=\frac{x}{x^c}$, and therefore 
independent of $k$.  Note that $\tilde{G}_{\!p^c}$ and $\tilde{G}_{\!x^c}$ are 
exactly the invariant vector fields of the manifold of $(p^c,x^c,\theta)$ 
corresponding to the contracted symmetry from the Heisenberg-Weyl group. To 
summarize, we have the set of $\tilde{G}_{\!\omega}^c$, $\tilde{G}_{\!t}^c$, 
$\tilde{G}_{\!p^{c}}^c$, $\tilde{G}_{\!-x^{c}}^c$,  and $\tilde{G}_{\!\theta}^c$ 
giving the commutators exactly as the as the old set of $\tilde{G}_{\!\omega}$, 
$\tilde{G}_{\!t}$,  $\tilde{G}_{\!p}$, $\tilde{G}_{\!-x}$,  and 
$\tilde{G}_{\!\theta}$, with the factors of 2 all replaced by $\hbar$.  With  
$\tilde{G}_{\!\theta}$ taken out, the rest constitute a representation of the 
contracted Galilean symmetry without the $U(1)$ central extension, which 
at the abstract Lie algebra level is trivialized and decoupled from the rest. 

Next, take the multiplicative operators $G_{\!\omega^{ij}}^c=x_i^c p_j^c - x_j^c p_i^c$,
${G}_{\!t}^c=\frac{ p_i^c  p^{ic}}{m}$,  ${G}_{\!p^{c}}^c=x^c$,
and ${G}_{\!-x^{c}}^c= p^c$. We have the formal relation 
\bea &&
G_{\!\omega^{ij}}^c = \frac{\hbar}{k^2} G_{\!\omega^{ij}} \;,
\qquad\qquad
{G}_{\!t}^c =  \frac{\hbar}{k^2} {G}_{\!t}  \;,
\nonumber \\ &&
{G}_{\!p^{c}}^c = \frac{\sqrt\hbar}{k} G_p \;,
\qquad\qquad
{G}_{\!-x^{c}}^c = \frac{\sqrt\hbar}{k} G_{-x} \;.
\eea
The commutator results for the classical operators with the $\tilde{G}^c_{\!s^c}$ 
( with $\omega^c=\omega$ and $t^c=t$) set above correspond again to results 
in Eq.(\ref{GtG}) with $2$ replaced $\hbar$. Thus, we recover the full algebraic 
structure introduced in Ref.\cite{L} for the Koopman-von Neumann classical setting.

\section{To the  Koopman-von Neumann  Classical Dynamics}
Finally, we check the explicit dynamical description obtained for the
classic setting, focusing especially on the Koopman-von Neumann  
formulation. The Schr\"odinger equation, Heisenberg equation,
and Liouville equation are to be cast in the following forms in the contraction limit 
\bea
i\hbar \frac{d}{dt} \phi(p^c,x^c;t) = \frac{k^2}{2} \kappa^c(p^c,x^c) {\star^c} \;\phi(p^c,x^c;t) 
   \to \infty \;,
\eea
\bea \label{Heomk}
\frac{d}{dt} \za(p^c,x^c;t) = \frac{k^2}{2i\hbar}\{\za(p^c,x^c;t), \kappa^c(p^c,x^c) \}_{\star^c} 
\to \{ \za(p^c,x^c;t),\kappa^c(p^c,x^c) \}\;,
\eea
\bea
 \frac{d}{dt} \rho(p^c,x^c;t) =\frac{k^2}{2i\hbar} \{ \kappa^c(p^c,x^c), \rho(p^c,x^c;t)  \}_{\star^c} 
\to \{ \kappa^c(p^c,x^c),\rho(p^c,x^c;t)  \}\;,
\eea
with the classical (antisymmetric) Poisson bracket $\{\cdot, \cdot\}$ of classical
phase space coordinates $(p^c,x^c)$. So, the Schr\"odinger equation on the
Hilbert space of pure state fails to make sense at the contraction limit, while
the Heisenberg equation and the Liouville equation give the correct classical
results.  The problem of the Schr\"odinger equation is not beyond one expectations. 
The Hilbert space of pure states, as an irreducible unitary representation,
collapses to the simple sum of one-dimensional subspaces of the coherent
states, so there is no continuous evolution to described on it any more.  Recall
that the Heisenberg equation can be seen as one on $\tilde{\mathcal{K}}$,
and hence it survives. Moreover, the Liouville equation is the Schr\"odinger equation
on $\tilde{\mathcal{K}}$. The reducible representation is on the Hilbert
space $\tilde{\mathcal{K}}$ containing all the states -- pure or mixed -- and therefore it is
not at all bothered by the fact that most of the pure quantum states become 
mixed states in the classical limit. Furthermore, note that the (classical) Liouville 
equation is insensitive to the rescaling/renormalization of $\rho$ to 
$\rho^c$, and similarly for the classical equation of motion going from
$\za$ to $\za^c$.

In the Koopman-von Neumann formulation, a classical wavefunction $\phi^c$
is to be introduced with $\rho^c\equiv|\phi^c|^2$ for each $\rho^c$. 
Each $\phi^c$ describes a mixed state in general, as does $\rho^c$. The 
Koopman-von Neumann Hilbert space is one of a reducible representation,
A Koopman-Schr\"odinger equation for $\phi^c$ relating to the classical
Liouville equation $\frac{d}{dt} \rho^c=  \{\kappa^c, \rho^c \}$ can be 
written then as
\bea
i\hbar  \frac{d}{dt} \phi^c = \kappa^c \phi^c \;.
\eea
One can rewrite the classical equation of motion in the Koopman-Heisenberg  
form \cite{cHsp} as 
\bea
\frac{\partial}{\partial t}  M_\za    =M_{\{ {\kappa^c}, \za \}}=[{X}_{\kappa^c}, M_\za] \;,
\eea
where ${X}_\kappa\!=\!\!\left[ \frac{\partial \kappa}{\partial p^c_i}\frac{\partial}{\partial x^{ic}}
-\frac{\partial \kappa}{\partial x^{ic}}\frac{\partial}{\partial p^c_i}\right]$ 
is the (classical) Hamiltonian vector field, which gives 
\bea \label{KH}
M_{\za(t^)} =e^{it(-i{X}_{\kappa^c})} M_\za \, e^{-it(-i{X}_{\kappa^c})} 
=e^{t{X}_{\kappa^c}} M_\za e^{-t{X}_{\kappa^c}} 
\;.
\eea
Recall that the multiplicative operator $M_\za=\za$ is just the classical limit 
of the $\za\star$ [{\em cf.} Eq(\ref{M})]; hence it is a simple multiplication with 
$\za (p^c,x^c)$ on the classical Hilbert space $L^2(\Pi)$. Taking a closer look, 
we see that the solution to the equation of motion (\ref{Heomk}) before taking
$k\to \infty$ can be written as
\bea\label{sHeom}
\za(t)\star^c=e^{\frac{k^2}{2\hbar}{it}(\kappa^c\star^c)} [\za(0)\star^c] 
      e^{-\frac{k^2}{2\hbar}{it}(\kappa^c\star^c)}\;.
\eea
This equation is nothing other than the $\za(t)\star$ solution to the original
Heisenberg equation of (\ref{heK}) written in terms of the rescaled classical 
variables. Expanding Eq.(\ref{starc}) and keeping only the first two terms, we have 
\bea
U_{\!\star}(t) = e^{-\frac{k^2}{2\hbar}{it}(\kappa^c\star^c)} \rightarrow 
   e^{-\frac{k^2}{2\hbar}{it} \left(\kappa^c-\frac{2i\hbar}{k^2} {X}_\kappa^c\right)}
   =e^{-\frac{ik^2t}{2\hbar}\kappa^c} \, e^{- t{X}_\kappa^c} \;.
\eea
This result is obviously consistent with Eq.(\ref{KH}), as the first exponential 
factor simply cancels itself out.  The classical limit is taken as the $k\to\infty$ limit,
but the dynamics is determined by the noncommutative part of the star product; 
therefore it is determined by the first nontrivial term in the expansion, which is also 
the dominating real term. For the Schr\"odinger picture considerations, however, 
one would keep only the dominating first term. The limit $U_{\!\star}(t)$ is then
consistent with the limiting  Schr\"odinger equation, but both involve
the diverging $k^2$ factor.  Again, the quantum Schr\"odinger equation 
is an equation of motion for the pure states the classical limit, and there
do not form a connected set in the reduced Hilbert space (except formally
at the zero magnitude point). The Koopman-Schr\"odinger equation is
exactly given by putting $k^2=2$ back into the limiting  Schr\"odinger equation 
for the diverging $k$.

The solution to the Koopman-Schr\"odinger equation is given in terms of
the Koopman-Schr\"odinger flow 
$U^{\ssc KS}(t)=e^{\frac{-it}{\hbar} \mathcal{G}_{\!\kappa}}$
in Ref. \cite{cHsp} with the generator $\mathcal{G}_{\!\kappa}$ given by
\bea
\mathcal{G}_{\!\kappa} = M_\kappa + M_{\!\vartheta(\kappa)} -i\hbar X_\kappa \;.
\eea
The first two terms contribute a change of a complex phase for $\phi^c$ with
no effect in the Heisenberg picture. The last term, and thus the whole set of
$U^{\ssc KS}(t)$, gives the Koopman-Heisenberg equation we obtained above,
as well as a time translation of (the magnitude of) $\phi^c$ in the 
Schr\"odinger picture.  The $M_{\!\vartheta(\kappa)}$ part is responsible 
for the geometric phase \cite{gp1,gp2}, a notion which requires formulating 
states, quantum or classical, as sections of a $U(1)$ principal bundle or 
a Hermitian line bundle \cite{gq,gp1,gp2} for its description. 
$M_{\!\vartheta(\za)}\! -\!i\hbar X_\za$ is really a covariant derivative 
($\vartheta$ a connection form) associated to the function $\za(p^c,x^c)$ 
which guarantees 
$\mathcal{G}_{\!\{\za,\zb\}}=i\hbar[\mathcal{G}_{\!\za}, \mathcal{G}_{\!\zb}]$,
{\em i.e.} the operators $\mathcal{G}_{\!\za}$ form a representation of same
Lie algebra as the Poisson algebra. Adopting the canonical trivialization of 
the $U(1)$ bundle over $\Pi$, coordinated by  $p^{c}_i$-$dx^{ic}$ as a 
K\"ahler manifold with a Euclidean metric on  
($d\vartheta=dx^{ic}\wedge dp^{c}_i$ is the symplectic form)
\footnote{The metric is essentially the restriction of the Fubini-Study metric 
on the quantum phase space (the projective Hilbert space) as the K\"ahler 
manifold $\mathds{CP}^\infty$ to the coherent state submanifold \cite{BZ}. 
It is hence totally compatible with the quantum description.
}, 
$\mathcal{G}_{\!\za}$ can be taken as acting on the wavefunction 
$\phi^c$ $\in L^2(\Pi)$ with $\vartheta(\za)\!=\!\! -\frac{1}{2} \left[
   p^c_i \frac{\partial \kappa}{\partial p^c_i}
 + x^c_i\frac{\partial \kappa}{\partial x^c_i}  \right]$ \cite{T}.  
It would be interesting to see a full $U(1)$ bundle formulation of the
WWGM formalism and its contraction limit, which is however beyond
the scope of this article.

\section{Conclusions}
We have explicitly presented a version of the WWGM formalism for 
quantum mechanics, which we propose as the most natural prescription 
unifying,  the formalism most familiar to a general physicist (the one base
on a Hilbert space of wavefunctions) and the abstract mathematical 
algebraic formalism related to noncommutative geometry. On the 
(pure state) Hilbert space ${\mathcal K}$ of wavefunctions $\phi(p_i,x_i)$
from the canonical coherent state basis, the observable algebra as a functional 
algebra of the $P_i$ and $X_i$ operators $C(P_i,X_i)$ can be seen as both the 
operator (functional) algebra $C(p_i\star,x_i\star)$ as well as $C(p_i,x_i)$ with 
a Moyal star product; $\za(p_i\star,x_i\star) \phi = \za(p_i,x_i)\!\star \phi$.
We advocate the former picture and the important notion that the algebra
is essentially an irreducible (cyclic) representation of the group ($C^*$-)algebra
from the relativity symmetry  within which the Hilbert space is a representation
for the group. The modern mathematics of noncommutative geometry \cite{P1,P2,C} 
essentially says that the noncommutative algebra $C(p_i\star,x_i\star)$ is to be 
seen as an algebra of continuous functions of a geometric/topological space
with the six noncommutative coordinates $p_i\star$ and $x_i\star$, and 
coordinates are of course the basic observables in terms of which all other 
observables can be constructed. $C(p_i\star,x_i\star)$ as a $C^*$-algebra
corresponds to the set of compact operators on ${\mathcal K}$ is a Moyal
subalgebra of $I\!\!B({\mathcal K})$ as given by ${\mathcal M}'$ (which is
a $W^*$-algebra). The mathematics also offers another geometric object as
a kind of dual object to the $C^*$-algebra, namely the space of pure states 
$\omega_\phi$ \cite{S}, which is equivalent to the (projective) Hilbert 
space (of ${\mathcal K}$) \cite{kS}. The projective Hilbert space is the
infinite-dimensional K\"ahler manifold $\mathds{CP}^\infty$ \cite{BZ,km}, 
with a set of `six times $\infty$' homogeneous coordinates. One key purpose 
of the article is to help a general physicist to appreciate such a perspective. 
Of course such an algebraic-geometric perspective also works perfectly well 
with Newtonian physics for which the observable algebra is commutative and 
contains functions of the classical phase space coordinates. We illustrate 
here also how that classical limit is retrieved from the quantum one.

This geometric notion is usually considered as only about the quantum 
phase space. Actually, the standard description of quantum mechanics breaks 
the conceptual connection between the phase space, the configuration space, 
and physical space itself in classical mechanics -- physical space is the 
configuration space (all possible positions) of a free particle, or of the 
center of mass as a degree of freedom for a closed system of particles; 
the configuration space is sort of like half the phase space, with the other 
half being the momentum space of conjugate variables. However, from 
both the noncommutative geometry picture and the $\mathds{CP}^\infty$ 
picture (for a single quantum particle) discussed above, it certainly looks 
like it does not have to be the case. In particular, $\mathds{CP}^\infty$ 
is a symplectic manifold and the Schr\"odinger equation is an infinite 
set of Hamiltonian equations of motion with the configuration and conjugate 
momentum variables taken as, say, the real and imaginary parts of $\phi(p_i,x_i)$ 
at each $(p_i,x_i)$ (or those of $\lla\phi|a_n\rra$ for any set $\left|a_n\rra$ 
of orthonormal basis). In Ref.\cite{066}, we have constructed a quantum model 
of physical space, or the position/configuration space of a particle, along 
parallel lines of the coherent state phase space construction as a representation 
of the relativity symmetry. Moreover, we showed that the model reduces back to 
the Newtonian model as the classical limit formulated as a relativity symmetry 
contraction limit. Part of the analysis in the current article was motivated by 
the idea of illustrating the solid dynamical picture underlying that framework. 

The quantum physical space obtained in Ref.\cite{066} is actually a Hilbert 
space equivalent to that of the phase space. The key reason is that for the 
quantum relativity symmetry $\tilde{G}(3)$ as the $U(1)$ central extension of 
the classical Galilean symmetry $G(3)$, phase space representations are generally 
irreducible while in the classical case they may be reduced to a sum of the 
position/configuration space and the momentum space ones. The central charge 
generator, as the $X_i$-$P_i$ commutator generates, a complex phase rotation 
in relation to the natural complex structure in $X_i +i P_i$ with the complex 
coordinates $\phi(p_i,x_i)$ mixing the position/configuration coordinates with 
the momentum coordinates, the division of which would otherwise be respected 
by the other relativity symmetry transformations. The analysis here establishes
explicitly that the $\tilde{G}(3)$ group plays the full role of a relativity
symmetry for quantum mechanics with the quantum model of the 
physical space and gives all the corresponding aspects for the Newtonian
theory as an approximation to be described as a relativity symmetry
contraction. The results also gives a comprehensive treatment of the
classical limit of quantum mechanics, to which there are otherwise
quite some confusing notions about in the literature.

The explicit analysis in Ref.\cite{066} focused only on the $H_{\!\ssc R}(3)$ 
subgroup with the time translation generator taken out, which is good 
enough for the mostly kinematical considerations there. Along these lines, we
put strong emphasis on the (relativity) symmetry group as the starting point.
The observable algebra is essentially the group ($C^*$) algebra or an irreducible
representation of it. Actually, we focus only on the Heisenberg-Weyl 
subgroup $H(3)$, and take into consideration the full relativity symmetry 
$\tilde{G}(3)$ only as unitary transformation on the Hilbert space and
as automorphisms on the observable algebra. All of this works very well because 
the relevant (e.g. spin zero) representation of the $\tilde{G}(3)$ group algebra 
is contained in $C(p_i\star,x_i\star)$. This is a natural parallel to the Hilbert
space of pure states as an representation of $H(3)$ and $H_{\!\ssc R}(3)$
(or $\tilde{G}(3)$). This is more or less the physical statement
that (orbital) angular momentum and Hamiltonian variables/operators
are to be written in terms of the position and momentum ones. It is really
a consequence of the structure of $\tilde{G}(3)$ with the series of
invariant subgroups
\[
U(1) \prec H(3)  \prec H_{\!\ssc R}(3)   \prec   \tilde{G}(3) \;,
\]
giving the following semidirect product structures:
\[
\tilde{G}(3)=H(3) \rtimes (SO(3)\times T )\;,
\]
where $T$ denotes the one parameter group of time translations.
The other relativity transformations act on $H(3)$ as outer
automorphisms and on its group algebra as inner automorphisms. 
Again, the Hilbert space as a group representation naturally sits inside the
representation of the group algebra with the natural (noncommutative, 
algebraic) multiplicative actions of the latter as the operator actions.

It is also interesting that while the rotational symmetry $SO(3)$ is
naturally to be included in the mathematical picture of even just the
$H(3)$ symmetry, the Galilean time translation is not. Moreover,
we have no problem describing the 
transformations generated by any real/Hermitian Hamiltonian
function $\kappa(p_i,x_i)$ or operator  $\kappa(p_i,x_i)\star$ as unitary 
transformations on the Hilbert space and automorphisms of the observable 
algebra, just like any Hamiltonian flow on a symplectic manifold. But then 
there is no reason to single out the parameter of a particular Hamiltonian
flow as physical time and the generator of physical energy. We may 
have to look for a more natural relativity symmetry framework in order 
to truly understood time, for example with Lorentz symmetry incorporated. 
For the current authors, we are particularly interested in using the relativity
symmetry as the basic key mathematical structure, and plan on pushing
forward for models of quantum spacetime and its related dynamics on 
the deep microscopic scale based on the idea of relativity symmetry
deformation/stabilization \cite{030}.

In summary, quantum mechanics can be, and we believe should be, 
seen as a theory of particle dynamics on a quantum/noncommutative 
model of the physical space with a picture as the infinite dimensional
K\"ahler manifold $\mathds{CP}^\infty$. It has a relativity 
symmetry of $\tilde{G}(3)$ and the observable algebra is naturally
the representation of the group $C^*$-algebra corresponding
to the representation of  $\tilde{G}(3)$ (time-independent
spin-zero) that describes the physical space.  The WWGM
formalism is just such a representation theory, and hence also
essentially the Hilbert space theory. Dynamics is included in the 
Hamiltonian flows on $\mathds{CP}^\infty$ as well as the 
corresponding automorphism flows on the $C^*$-algebra. 
The mathematical framework is valid for any group as relativity
symmetry, and a group obtained as the contraction limit of
another serves as an approximation of the latter with the full
theory retrievable from pushing the contraction throughout the
original theory, as our illustration of obtaining the Newtonian 
theory from quantum mechanics. Lie group/algebra deformations
in the reverse process to contraction, hence giving natural
candidates for theories the quantum and classical mechanics
serves as approximation. The fully deformed/stabilized 
(special) relativity symmetry, probably for Planckian physics,
is expected to give full noncommutativity among all $X$ and $P$
to which quantum mechanics is the minimal case with
noncommutativity. It suggests all noncommutative models
of spacetime should have be phase space models; 
energy-momentum is much a part of the physical space 
only in the classical approximation to which one can consider
the configuration and the momentum parts separately.

\vspace*{.2in}
\noindent{\bf Acknowledgements \ }
The authors are partially supported by research grant 
MOST 105-2112-M-008-017 and MOST 106-2112-M-008-008 from the 
MOST of Taiwan.


\end{document}